\documentclass[onecolumn]{autart}    

\usepackage{graphicx}          
\usepackage{amsmath}
\usepackage{url}
\usepackage{array,tabularx}
\usepackage{graphics,color} 
\usepackage{epsfig} 
\usepackage{mathptmx} 
\usepackage{times} 
\usepackage{amssymb}  
\usepackage{xfrac}
\usepackage{comment}
\usepackage{gensymb}
\usepackage[export]{adjustbox}

\makeatletter
\DeclareRobustCommand{\qed}{%
  \ifmmode 
  \else \leavevmode\unskip\penalty9999 \hbox{}\nobreak\hfill
  \fi
  \quad\hbox{\qedsymbol}}
\newcommand{\openbox}{\leavevmode
  \hbox to.77778em{%
  \hfil\vrule
  \vbox to.675em{\hrule width.6em\vfil\hrule}%
  \vrule\hfil}}
\newcommand{\qedsymbol}{\openbox}

\newcommand{\proofname}{Proof}
\makeatother

\begin{document}

\begin{frontmatter}

\title{Real-time-controlled artificial quiet channel for acoustic cloaking under varying detection conditions}

\author[Jerusalem]{Or Lasri} \: and \:
\author[Jerusalem]{Lea Sirota}\ead{leabeilkin@tauex.tau.ac.il}

\address[Jerusalem]{School of Mechanical Engineering, Tel Aviv University, Tel Aviv 69978, Israel}

\begin{abstract}

We consider the problem of hiding non-stationary objects from acoustic detection in a two-dimensional environment, where both the object's impedance and the properties of the detection signal may vary during operation.
The detection signal is assumed to be an acoustic beam created by an array of emitters, which scans the area at different angles and different frequencies.
We propose an active control-based solution that creates an effective moving dead zone around the object, and results in an artificial quiet channel for the object to pass through undetected. 
The control principle is based on mid-domain generation of near uni-directional beams using only monopole actuators. 
Based on real-time response prediction, 
these beams open and close the dead zone with a minimal perturbation backwards, which is crucial due to detector observers being located on both sides of the object's route.
The back action wave determines the cloak efficiency, and is traded-off with the control effort; the higher is the effort the quieter is the cloaking channel.
We validate our control algorithm via numerical experiments in a two-dimensional acoustic waveguide, testing variation in frequency and incidence angle of the detection source. Our cloak successfully intercepts the source by steering the control beams and adjusting their wavelength accordingly.

\end{abstract}

\end{frontmatter}

\section{Introduction}    \label{Intro}

Acoustic cloaking can be regarded as the use of devices, materials, actions, or their combination to prevent acoustic detection of an object.
The acoustic detection process is based on capturing sound fields indicating the existence of the object. These sound fields can originate either from self-emission, which is dubbed the acoustic signature, or from external sources that emit sound waves towards the object and record the back-scattered field. 
Despite numerous solutions that have been suggested over the years, acoustic cloaking remains one of the most exciting and intriguing problems in Engineering. This is partially due to the endless setups, configurations, operating conditions and constraints of the objects to be cloaked, as well as of the associated detection conditions, each posing its own challenge and requiring its own targeted solution. \\
The cloaking problem can be formulated in different ways.
One is scattering suppression and/or absorption.
For this formulation, a common approach includes passive shells covering the object \cite{sohrabi2020stochastic,fu2021review}. In a more advanced version these shells are given by architectured structures, also known as metamaterials, which are artificially designed to realize, through the collective dynamical behavior of their unit cells, properties that are unavailable in natural materials. 
In particular, patterns of foams and metal plates cut into labyrinthine units, perforated with holes or machined into cavities, intricately layered structures, and many other sophisticated designs were suggested \cite{li2016acoustic,zhang2016three,tang2017deep,tang2017hybrid,wang2018space,zhu2019multifunctional,li2021ultra}.
Other types of metamaterial-based solutions, which originated in electromagnetic systems and were adapted in acoustics, are the transformation cloak \cite{pendry2006controlling,cummer2007one}, in which waves are guided around the object without far-field distortion, the carpet cloak \cite{li2008hiding,popa2011experimental,zigoneanu2014three}, in which objects laying on the ground are concealed by properly shifting the detection wave phase imitating bare ground, the mantle cloak \cite{alu2009mantle}, and more \cite{cummer2016controlling}. \\
What makes the passive design advantageous is first of all the ease of fabrication, for example by machining or printing, which enables adding as many unit cells as required. Another advantage is the ease of use, requiring just covering the object with the shell. 
There are several situations, however, in which passive solutions become less effective. One situation is related to low frequency, hence long wavelength detection signals. The longer are the wavelengths, the ticker and thus the heavier the absorbing coatings usually become (e.g. a low frequency anechoic chamber \cite{weitsman2020effects}).
Multiplying by the coverage area, especially for large-scale objects, the passive shell weight may grow impractically high. %
Although subwavelength and thus thin coatings that can efficiently absorb low frequency signals do exist \cite{li2016acoustic,xie2017coding}, they are usually effective in a particular frequency range. This brings us to the second situation, which is inconsistency in detection conditions. \\
The detection signal frequencies, the relative position between the object and the detector, as well as the object's own impedance
may vary during operation. 
To address these situations, tunable designs have been suggested \cite{fan2020reconfigurable,zhou2020tunable}, mostly based on adjustable geometry of the unit cells. Such designs significantly expand the range of the cloaking effectiveness, while keeping the shell thin and lightweight.
If the tunability is enabled only offline, the cloak is efficient when the change in the detection conditions is known in advance. 
However, if the detection conditions vary in real-time, online tunability becomes necessary. 
Active structures, which are operated by external energy sources and are
encountered in diverse waveguiding applications \cite{zangeneh2019active,zhou2020voltage,chen2021reprogrammable,geib2021tunable,popa2022new}, could then be utilized.
For example, a type of cloak that employs active real-time modulation is the spacetime cloak \cite{mccall2010spacetime,kinsler2014cloaks}, the goal of which is hiding acoustic events of finite duration. In that cloak the medium dispersion is manipulated in time rather than in space, creating a 'hole' in time. \\
In this work, we consider the situation in which the cloak properties need to be based on actual real-time measurements of the detection field and the object's response. 
Measurement-based active designs usually include a host structure and external actuators at the degrees of freedom. 
The actuators generate inputs based on commands of embedded electronic controllers, which process the system's dynamical response in real-time. 
The controllers can be reprogrammed at will by the user, implying that the overall structural dynamics is determined by the program and not by a particular element, passive or active. 
The resulting capabilities include real-time tuning of existing properties, creating long-range interactions via distant site measurements, creating new, possibly non-physical dynamical behaviors of any kind, 
and switching between all these functionalities on the same platform, without the need to refabricate the elements. 
In addition, since the main components of such structures are actuators and sensors, which in acoustics are usually given by flat surfaces, their total weight can be significantly smaller than the weight of passive shells. \\
Actively-controlled measurement-based approach has recently emerged in diverse electromagnetic, acoustic and elastic wave manipulation applications \cite{baz2010active,popa2015active,becker2018immersive,hofmann2019chiral,brandenbourger2019non,darabi2020experimental,scheibner2020non,rosa2020dynamics,cho2020digitally,ghatak2020observation,sirota2020non,sirota2021real,sirota2021quantum,kotwal2021active,fruchart2021non,you2021reprogrammable,jalvsic2023active}, enabling exotic wave dynamics, such as violation of Newtonian mechanics, non-reciprocal propagation, adaptive refocusing, or artificial boundary conditions for simulation domain scaling. 
Utilizing this approach, we design an actively-controlled acoustic cloak in a form of an artificial quiet channel, which creates a moving dead zone around the object to be concealed, and adapts in real-time to changes in certain detection properties. 
In Sec. \ref{Setup} we describe our cloaking system setup with assumed constraints on actuation and measurement locations, as well as the detection system setup and its parameters uncertainties. 
In Sec. \ref{Algorithm} we derive the control algorithm, which is based on a technique we term as near unidirectional wave generation, necessary for the artificial dead zone creation.
In Sec. \ref{Time_sim} we simulate the performance of our cloak, and demonstrate its effectiveness for different frequencies and incidence angles of the detection beam. The work is discussed and summarized in Sec. \ref{Conclusion}.

\section{Cloaking system setup}  \label{Setup}

\begin{figure}[tb]
\begin{center}
\begin{tabular}{l}
 \textbf{(a)}    \\ \includegraphics[height=4.5cm, valign=t]{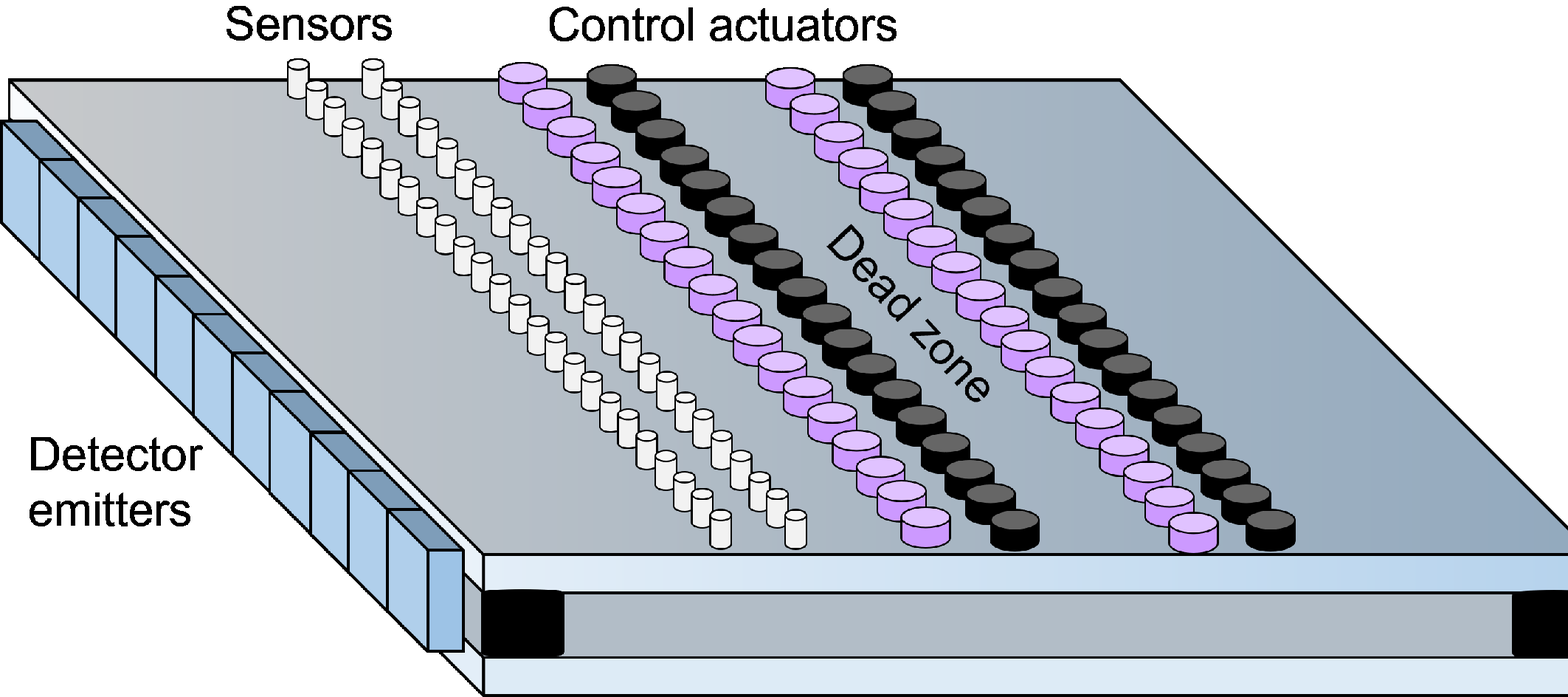}
 \end{tabular} 
 \begin{tabular}{l l l}
 \textbf{(b)} & & \textbf{(c)} \\
 \includegraphics[height=7.3cm, valign=c]{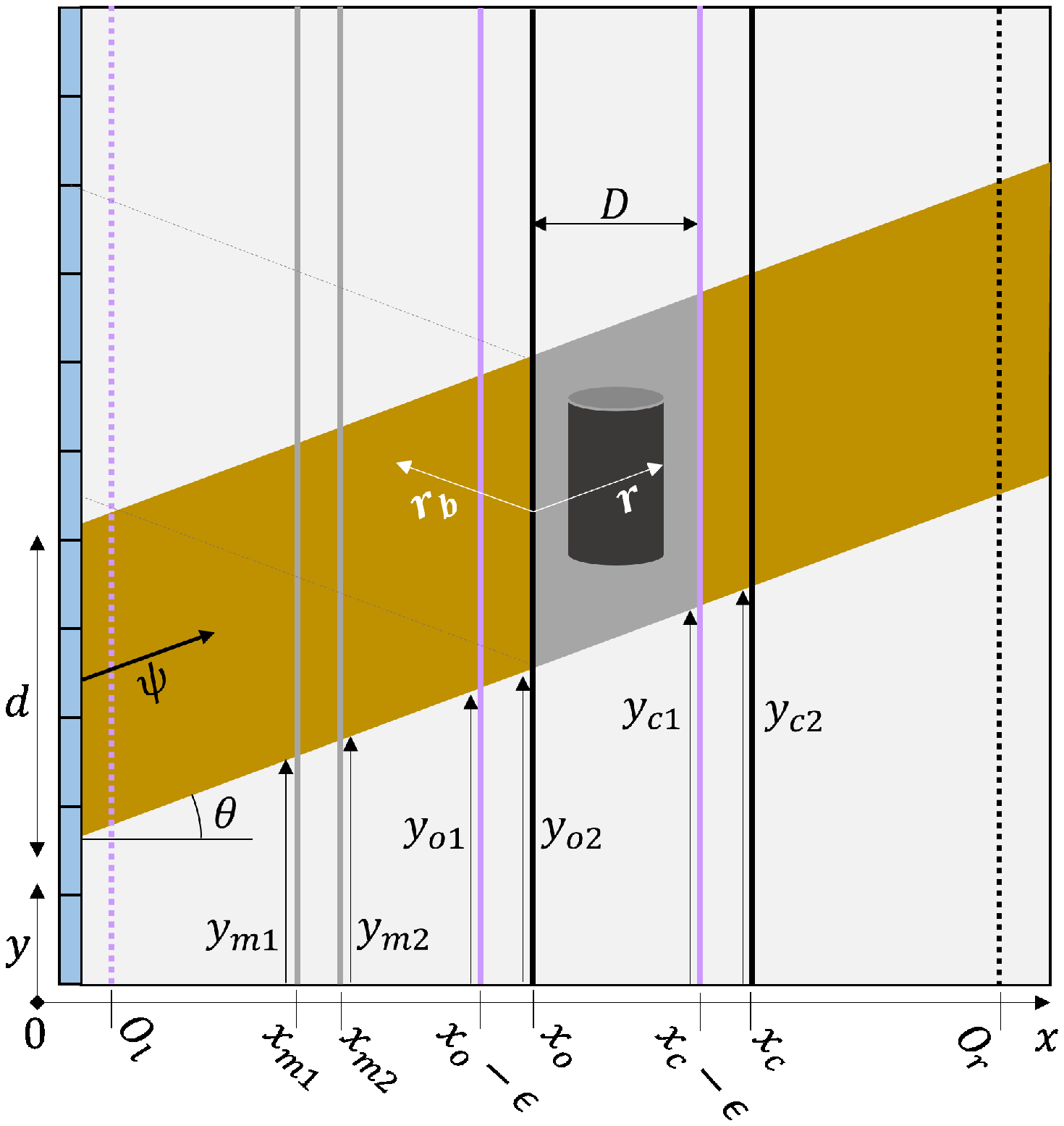} & & \includegraphics[height=5.7cm, valign=c]{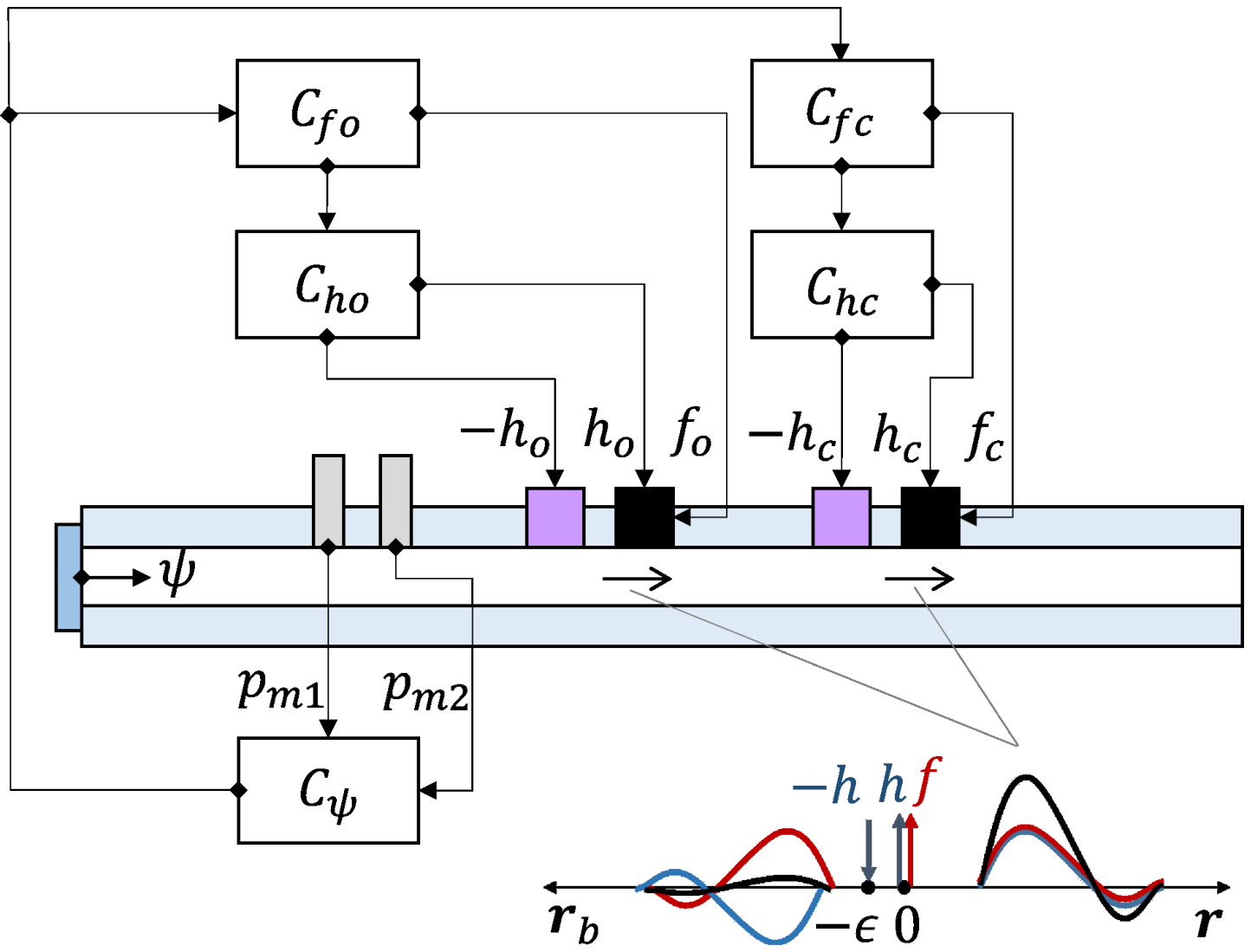}
\end{tabular}
\end{center}
\caption{Artificial quiet channel cloak. (a) The model schematic. The underlying platform is a two-dimensional waveguide consisting of gapped parallel plates, which supports sound propagation between the plates. The sound pressure field is created by a series of emitters, blue cuboids, along one of the boundaries, constituting the detectors. An active cloaking system is implemented in the waveguide using two pairs of control actuator arrays, black and purple cylinders, which create an artificial dead zone between the arrays. The control operation is based on real-time pressure field measurements performed by two sensor arrays, gray cylinders. (b) The working principle of the cloak. Depicted is the horizontal cross-section between the plates. The detection sources emit a sound beam $\psi$, gold strip, of frequency $\omega$ and of $y$ extension $d$, which scans the area at an angle $\theta$. $d$, $\omega$ and $\theta$ are unknown and may vary. The actuators at $(x_o,y_{o2})$ and $(x_o-\epsilon,y_{o1})$ ($(x_c,y_{c2})$ and $(x_c-\epsilon,y_{c1})$) generate a near uni-directional control beam in the $\textbf{r}$ direction with a backwave in the $\textbf{r}_b$ direction, which opens (closes) the dead zone, gray shadowed area. The object, dark gray cylinder, regardless of its possibly varying impedance, remains undetected by observers along $O_r$, dotted-black, and $O_l$, dotted purple. (c) The control scheme, depicted at the side view cross-section. 
The controller $C_\psi$ predicts the detection wave response along actuation arrays, based on real-time information about the properties of $\psi$, assumed to be obtained from measurements along $x_{m1}$ and $x_{m2}$.
Controllers $C_{ho}$ and $C_{hc}$ create the total unidirectional opening and closing waves $f_o$ and $f_c$ out of the pairs $f_o,h_o$ and $f_c,h_c$. Controllers $C_{fo}$ and $C_{fc}$ generate the commands for $f_o$ and $f_c$. Inset: illustration of the near uni-directional wave formation in the $\textbf{r}$ direction with a backwave in the $\textbf{r}_b$ direction.}
\label{GenScheme}
\end{figure}

Our goal is to 
prevent the detection of objects 
in a two-dimensional nondispersive medium, where the properties of the object, such as its impedance, as well as the geometrical and spectral properties of the detection signal, are not known in advance and can vary during the detection sessions. 
To handle these variations we design a cloaking system that is based on active real-time control. 
Our representative platform is a waveguide consisting of two rigid parallel plates with a gap between them, as illustrated in the schematic of Fig. \ref{GenScheme}(a). For a gap sufficiently small with respect to the wavelength, the propagation in the gap can be approximated as two-dimensional. \\
In addition, to narrow down this very general problem, as well as to account for constraints that may arise in practical situations, we consider the following set of conditions:
(i) There is only one detection emitters array, located either to the left or to the right of the object.
The detection receivers array, on the other hand, is assumed to exist on both sides of the object. 
(ii) The detection signal is a single frequency harmonic beam of finite duration that scans the waveguide area. The beam spectral properties, including amplitude, frequency and phase shift, as well as its geometric properties, including scanning angle, spatial width and emitters distance from the object are unknown in advance. All these properties can vary in a discrete manner, i.e. the detection process takes place at finite time intervals, with breaks between the intervals. During a particular interval the properties are constant, but in the next interval they can be different.
(iii) The control actuators are acoustically transparent, and can be placed behind and in front of the object in an interior line. That is, they cannot be mounted on the object's boundary and cannot constitute a boundary themselves. The sensors can be placed only along an interior line as well.
(iv) The object's self emission is negligible. (v) The object cannot be covered by a rigid shell. \\
Following condition (i), we assume without loss of generality that the detection emitters are located along the left boundary of the waveguide. In Fig. \ref{GenScheme}(a) these emitters are sketched as blue cuboids. 
Condition (ii) enables us to treat the system as linear time-invariant between the intervals, thus facilitating control design, and yet reasonably accounting for the inevitable variation in detection properties.
To accommodate condition (iii), we consider monopole type actuators mounted in one of the plates, e.g. the upper one, and facing inwards, as illustrated by the black and purple cylinders in Fig. \ref{GenScheme}(a). The actuators are those that create the cloak. They are arranged in two pairs of arrays, which form and determine the cloaking region. The two arrays of gray cylinders indicate the sensors, which measure the sound pressure field between the plates. This particular two pair arrangement both of the actuators and the sensors is directly related to our control strategy, as described in Sec. \ref{Algorithm}. The actuation transducers are assumed to be nonresonating and not scattering the fields by their physical presence, which means they are acoustically transparent, as required. When these actuators are switched off, free space propagation between the plates is fully resumed.
Here, acoustic transparency essentially means that we cannot place actuators in the waveguide gap. \\
As for conditions (iv) and (v), the common approach is first to cover the actual object by a perfectly rigid shell, and then to design the cloak on top of it \cite{pendry2006controlling,cummer2007one}. That way the self emission of the object and its impedance do not matter at all. These solutions work well when the rigid coverage is possible, but since here it is not allowed, both self emission and self impedance become essential. Condition (iv) implies that only self impedance matters. Condition (v) then determines the key principle of our approach: \textit{canceling the detection wave in the propagation direction, in mid-domain, before it hits the object, i.e. before it becomes affected by the object's impedance, and not after that.} This is crucial for satisfying condition (v). \\
Our cloaking concept is illustrated in Fig. \ref{GenScheme}(b) in a top view of the medium between the plates.
The detection system launches a two-dimensional beam of $y$ extension $d$, amplitude $A$, frequency $\omega$ and phase shift $\phi$, which scans the area at angle $\theta$ in order to locate the object. 
The distance $D$ between the inner actuation arrays defines the region where the object (gray cylinder) can exist and needs to be cloaked. In this region the active cloaking system needs to suppress the detection beam, i.e to create a null intensity field that we denote by a dead zone, and then to reconstruct this field beyond this region so that it coincides with free space propagation. 
The distance between the actuator arrays within each pair is defined by $\epsilon$. The actuators along $x=x_o$ and $x=x_o-\epsilon$ are responsible for the beam suppression, i.e. for the dead zone opening, whereas the actuators along $x=x_c$ and $x=x_c-\epsilon$ are responsible for the reconstruction, i.e. the dead zone closing. This dead zone can be regarded artificial, since unlike passive absorption, here it comprises both the detection beam and the out-of-phase control beam. The control commands for all the actuators, including which of them should be activated (the locations $y_{o1},y_{o2}$ for opening and $y_{c1},y_{c2}$ for closing) and their required time response, are based on real-time prediction of the detection wave evolution along the actuators arrays. This prediction is based on real-time information about $d$, $A$, $\omega$, $\phi$ and $\theta$, which is assumed to be obtained from the sensors read-out along $x_{m1}$ and $x_{m2}$. \\
The general principle of the control algorithm is based on suppressing the detection beam at $x>x_o$ for all $y$, and then reconstructing it at $x>x_c$ as if the cloak did not exist. 
The cloak will be considered effective if the time response of the right and left observer arrays, respectively located along $x=O_r$, dotted black, and $x=O_l$, dotted purple, is close enough to an unperturbed detection wave.
The control wave, both at $x=x_o$ and $x=x_c$, thus needs to create as minimum backward disturbance as possible: at $x<x_o$ due to the $O_l$ observer, and at $x<x_c$ in order to keep the null intensity field in the cloaking region, which otherwise will cause reflection from the object and affect the $O_r$ observer. 
Therefore, the control beams need to be \textit{as uni-directional as possible} in the propagation direction $\textbf{r}$. \\
Uni-directional beam generation is most often associated with boundary sources, since then only the outside of the boundary region exists. In the context of wave control, boundary sources are used in active closed loop setups for wave suppression in the entire structure \cite{krstic2008output,sirota2015fractionalA,becker2018immersive,borsing2019cloaking,becker2021broadband}. However, launching such beams from a domain interior, i.e. when no physical boundaries present, is more intricate.
Theoretically, mid-domain uni-directional actuation can be achieved by launching a monopole and a dipole source simultaneously from the same interior point, as is sometimes done in numerical simulations. However, since this is not practical in general, and in particular in our case due to the restriction on placing actuators in the waveguide gap, our algorithm needs to generate a similar effect using a pair of monopole actuator arrays. This algorithm is presented next.

\section{The control algorithm}  \label{Algorithm}

To derive the algorithm, we first mathematically describe the field inside the waveguide and its coupling to the external sources.
The continuous acoustic medium of mass density $\rho_0$ and bulk modulus $b_0$ is defined by a scalar pressure field $p(x,y,t)$ and a vector flow velocity field $\textbf{v}(x,y,t)$, consisting of the components $v_x(x,y,t)$ and $v_y(x,y,t)$ in the $x$ and $y$ directions, respectively. Defining the detection signal by $\psi$, and the cloak opening and closing monopole arrays by $q_{o,1},q_{o,2}$ and $q_{c,1},q_{c,2}$, the field equations take the form
\begin{subequations}  \label{eq:Medium_eq}
\begin{align}
    \rho_0\frac{\partial \textbf{v}(x,y,t)}{\partial t}&=-\nabla p(x,y,t),   \label{eq:Medium_eq_M} \\ 
    \frac{1}{b_0}\frac{\partial p(x,y,t)}{\partial t}&=-\nabla\cdot \textbf{v}(x,y,t)+\psi(y,t)\delta(x)+\sum_{i=o,c}{ q_{i,1}(y,t)\delta(x-(x_i-\epsilon))+q_{i,2}(y,t)\delta(x-x_i)}. 
     \label{eq:Medium_eq_B}
\end{align}
\end{subequations}
The delta function $\delta(\cdot)$ indicates the position of the sources along the $x$ axis, so that $\psi$ is distributed along the edge $x=0$, $q_{o,1}$ and $q_{o,2}$ along $x=x_o$ and $x=x_o-\epsilon$, and $q_{c,1}$ and $q_{c,2}$ along $x=x_c$ and $x=x_c-\epsilon$.  
The overall control scheme is illustrated in Fig. \ref{GenScheme}(c), where we define
\begin{equation} \label{eq:decomposition}
    q_{o,1}=-h_o \quad, \quad q_{o,2}=h_o+f_o \quad , \quad q_{c,1}=-h_c \quad, \quad q_{c,2}=h_c+f_c.
\end{equation}
The control system begins with the controller $C_\psi$, which predicts the geometrical and spectral properties of the detection beam along $x=x_o$, $x=x_o-\epsilon$, $x=x_c$ and $x=x_c-\epsilon$.
Based on these properties, the commands for the control inputs $f_o$, $h_o$, $f_c$ and $h_c$ are produced. 
The controllers $C_{fo}$ and $C_{fc}$ directly relate $f_o$ and $f_c$ to the output of $C_\psi$. The controllers $C_{ho}$ and $C_{hc}$ respectively relate $h_o$ to $f_o$ and $h_c$ to $f_c$, and are completely independent of $C_{fo}$ and $C_{fc}$.
We begin with the design of $C_{ho}$ and $C_{hc}$, as detailed in Sec. \ref{uni}, and then proceed to the design of $C_\psi$, $C_{fo}$ and $C_{fc}$, as detailed in Sec. \ref{openclose}.

\subsection{Near uni-directional interior control wave generation}   \label{uni}

In this section we design the controllers $C_{ho}$ and $C_{hc}$, which represent the key principle of our artificial quiet channel cloak. 
The pairs $h_o,f_o$ and $h_c,f_c$ need to generate beams as uni-directional as possible in the $x,y$ plane along the angle $\theta$, which is the incidence angle of the detection beam $\psi(y,t)$. 
In Fig. \ref{GenScheme}(b) we defined this direction by $\textbf{r}$.
Inspired by recent ideas for a purely one-dimensional setup \cite{sirota2019active,sirota2020modeling}, but facing the complexity of the additional spatial dimension, we now derive an algorithm for the pair $h_o,f_o$ in two dimensions, where the same principle applies to the pair $h_c,f_c$.
If a coinciding monopole input $f_o(y,t)\delta(x-x_o)$ (with $h_o(y,t)=0$) and a hypothetical $\textbf{r}$ direction dipole input $d_o(y,t)$ were practically possible along the actuation line $x=x_o$, the term $D_o(x,y,t)=d_o(y,t)\delta(x-x_o)$ would be added to \eqref{eq:Medium_eq_M}.
Setting then $d_o(y,t)=z_0f_o(y,t)$, a perfectly unidirectional wave would be launched at $x>x_o$ in the $\textbf{r}$ direction, which is determined by the $y$ axis dependence of $f_o$. 
The monopole inputs $h_o(y,t)$ at $x_o$ and $-h_o(y,t)$ at $x_o-\epsilon$ thus need to mimic the dipole input $d_o(y,t)$. Their exact expression is obtained by combining \eqref{eq:Medium_eq_M} and \eqref{eq:Medium_eq_B} into a total second order wave equation, which requires differentiation of $D_o(x,y,t)$ with respect to $\textbf{r}$, and of $h_o(y,t)$ with respect to $t$, leading to
\begin{equation} \label{eq:eq_with_f_and_h}
    \frac{\partial D_o(x,y,t)}{\partial x}=z_0f_o(y,t)\lim_{r_\epsilon\rightarrow 0}\frac{1}{r_\epsilon}\left(\delta(x-x_o)-\delta(x-(x_o-\epsilon))\right)=\frac{\partial h_o(y,t)}{\partial t}\left(\delta(x-x_o)-\delta(x-(x_o-\epsilon))\right),
\end{equation}
where $r_\epsilon=\epsilon/\cos{\theta}$ is the direct distance between $h_o$ and $f_o$ in the $\textbf{r}$ direction, and the angle $\theta$ is expected from the measurements.
This implies that when $r_\epsilon$ is sufficiently small, $h_o(y,t)$ becomes the limit of a simultaneous differentiation of $f_o(y,t)$ in space with respect to $\textbf{r}$, and its integration in time,
\begin{equation}  \label{eq:h_o_exact}
    h_o(y,t)\rightarrow \frac{1}{\tau_{r\epsilon}}\int_0^t{f_o(y,t')\mathrm{d}t'}.
\end{equation}
Here, $\tau_{r\epsilon}=r_\epsilon/c$ is a time constant indicating the time required for a wave to travel the respective distance $r_\epsilon$, and $c=\sqrt{b_0/\rho_0}$ is the speed of sound in the medium. 
The relation in \eqref{eq:h_o_exact}, however, might lead to an unbounded $h_o(y,t)$ for $f_o(y,t)$ that contains zero frequency components. To define a stable control law relating $h_o$ and $f_o$ we consider an approximation of \eqref{eq:h_o_exact}, which in Laplace domain takes the form
\begin{equation}  \label{eq:C_h}
    h_o(y;s)=C_h(s)f_o(y;s) \quad , \quad C_h(s)=\frac{1}{\tau_{r\epsilon}(s+\eta)}
\end{equation}
for some constant $\eta>0$, which is a free design parameter. The controller $C_h$ in \eqref{eq:C_h} corresponds to both $C_{ho}$ and $C_{hc}$ in Fig. \ref{GenScheme}(c).
The distribution of $f_o(y,t)$ and $\pm h_o(y,t)$ along the $y$ axis is determined by a rectangular window of width $d$, modulated by a Gaussian envelope. The window begins at $y_{o1}$ for $\pm h_o(y,t)$ and at $y_{o2}$ for $f_o(y,t)$ ($y_{c1}$ and $y_{c2}$, respectively, for the $\pm h_c,f_c$ pair). 
The actual values of $y_{o1}$ and $y_{o2}$ ($y_{c1}$ and $y_{c2}$), as well as the $y$ axis dependence of $f_o$ ($f_c$) that is responsible for the rotation of the control beam at $\theta$, are determined by the algorithm of $C_{fo}$ ($C_{fc}$), as detailed in Sec. \ref{openclose}.
The control law \eqref{eq:C_h} generates a near unidirectional control beam in the $x>x_o$ half space.
The fact that the resulting control wave is near but not perfectly unidirectional is exhibited through a residual back action wave, generated in the $\textbf{r}_b$ direction, which we denote by a backwave.
The smaller is $\epsilon$ with regards to the propagating wavelength $\lambda$, the smaller is the backwave at $x<x_o$. %
The backwave depends also on the incidence angle. This can be illustrated by calculating the forward and backward responses in locations $r$ and $r_b$ along the $\textbf{r}$ and $\textbf{r}_b$ axes, respectively. 
The pressure field response $p_{f,h}$ to the control inputs $f_o$ and $\pm h_o$ can be written in Laplace domain as
\begin{equation}   \label{eq:G}
\begin{alignedat}{2}
    p_{\textrm{cont}}(r;s)&=\frac{1}{2}z_0e^{-\tau_r s}\left[f_o(y;s)+\left(1-e^{-\tau_{r\epsilon} s}\right)h_o(y;s)\right], \quad r\geq 0,&& \\
p_{\textrm{cont}}(r_b;s)&=\frac{1}{2}z_0e^{-\tau_{rb} s}\left[f_o(y;s)+\left(1-e^{+\tau_{r\epsilon} s}\right)h_o(y;s)\right], \quad r_b\geq 0,&&
\end{alignedat}
\end{equation}
where $\tau_r=r/c$ and $\tau_{rb}=r_b/c$ are the corresponding time constants. 
The control waves are given in \eqref{eq:G} in two regions of interest: in $r>0$, where cancellation of the detection wave is required, and in $r_b<-r_\epsilon$, where suppression of the residual control backwave is required. 
The transfer function from each concentrated input to the pressure $p_{\textrm{cont}}$ in \eqref{eq:G} is a scaled pure delay, which indicates that the output is a pure shift of the input. The propagating wave does not undergo any shape distortion, as expected for the non-dispersive system \eqref{eq:Medium_eq}.
With the control law \eqref{eq:C_h}, the response in \eqref{eq:G} takes the form
\begin{equation}  \label{eq:v_fh_lim}
\begin{alignedat}{2}
    p_{\textrm{cont}}(r;s)&=\frac{1}{2}z_0e^{-\tau_r s}Q(s)f_o(y;s), \quad r\geq 0,&& \\
p_{\textrm{cont}}(r_b;s)&=\frac{1}{2}z_0e^{-\tau_{rb} s}\overline{Q}(s)f_o(y;s), \quad r_b\geq 0,&&
\end{alignedat}
\end{equation}
where
\begin{equation}   \label{eq:Q}
\begin{alignedat}{2}
Q(s)&=1+A(s), \quad A(s)&&=\left(1-e^{-\tau_{r\epsilon} s}\right)C_h(s), \\
\overline{Q}(s)&=1+\overline{A}(s), \quad \overline{A}(s)&&=\left(1-e^{+\tau_{r\epsilon} s}\right)C_h(s).
\end{alignedat}
\end{equation}
As the distance $r_\epsilon$ and the free parameter $\eta$ become smaller, we obtain the limits
\begin{equation}  \label{eq:lim_eta}
\begin{alignedat}{2}
\lim_{r_\epsilon,\eta\rightarrow 0}A(s)&=1, \quad \lim_{r_\epsilon,\eta\rightarrow 0}\overline{A}(s)&&=-1, \\
\lim_{r_\epsilon,\eta\rightarrow 0}Q(s)&=2, \quad \lim_{r_\epsilon,\eta\rightarrow 0}\overline{Q}(s)&&=0.
\end{alignedat}
\end{equation}
When the limits in \eqref{eq:lim_eta} are approached, the control wave travels in the positive $\textbf{r}$ direction only, as indicated by \eqref{eq:v_fh_lim}, and illustrated by the inset in Fig. \ref{GenScheme}(c). 
The efficiency of the proposed near uni-directional mechanism is captured by the control effort that is required to achieve a particular reduction of the backwave.
The amplitude of the actual near uni-directional wave is determined by $Q(s)$. The amplitude of the backwave is determined by $\overline{Q}(s)$. 
The maximal effort of the control input $h_o$ increases when $\eta$ decreases and when the ratio $\lambda/\epsilon$ increases, where $\lambda$ is the beam wavelength. This constitutes a trade-off with the backwave amplitude, as demonstrated both in frequency domain and in time domain in Sec. \ref{Time_sim}, Fig. \ref{trade_off}.

\subsection{Response prediction at actuation locations, cloak opening and closing}    \label{openclose}

\begin{figure}[tb]
    \begin{center}    
     \includegraphics[height=8.2 cm, valign=c]{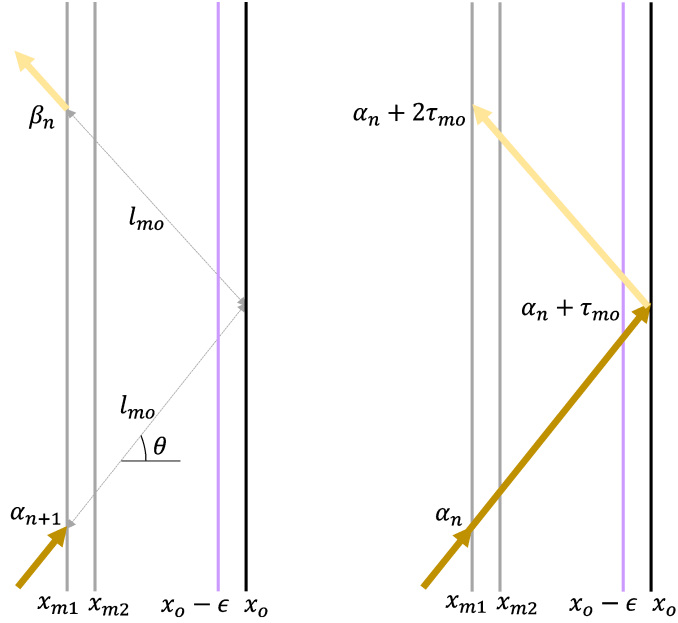} 
    \end{center}
    \caption{Progressive and regressive wave decoupling conditions. Left - the interval $n$ backwave needs to leave $x_{m1}$ no later than the arrival of the $n+1$ interval detection beam, implying the $\alpha_{n+1}\geq \beta_n$ requirement. Right - real-time parameters estimation should take place after the interval $n$ detection beam arrived at $x_{m1}$ but not later than the arrival of the $n$ interval backwave. This implies the $\alpha_{n+1}\geq \beta_n+2\tau_{mo}$ requirement, where $\tau_{mo}=l_{mo}/c$.}
    \label{fig:estimation}
\end{figure}

After designing the controller $C_h$ in \eqref{eq:C_h}, which relates $f_{o,c}$ with $h_{o,c}$, we now derive the algorithms for the resulting near unidirectional control waves, $f_o$ and $f_c$, which are respectively responsible for opening and closing the dead zone. 
These algorithms, respectively coded into the controllers $C_{fo}$ and $C_{fc}$, are based on real-time prediction of the detection beam free space evolution along the channel opening and closing locations, $x_o$, $x_o-\epsilon$, $x_c$ and $x_c-\epsilon$, carried out by the controller $C_\psi$. 
Following the finite intervals assumption (condition ii of Sec. \ref{Setup}), and the geometric definitions in Fig. \ref{GenScheme}(b), the detection beam as captured by the sensors along $x_{m1}$ and $x_{m2}$, is given by
a series of harmonic bursts at time intervals $\alpha_1\leq t \leq \beta_1$, $\alpha_2\leq t \leq \beta_2$, and so on, so that $\alpha_{n+1}\geq \beta_n$. At the $n_{th}$ interval the measured pressure field beams, $p_{m1}=p(x_{m1},y,t)$ and $p_{m2}=p(x_{m2},y,t)$, have an amplitude $A_n$, frequency $\omega_n$, and a respective phase shift $\phi_{m1\_n}$ and $\phi_{m2\_n}$. The beam has also the $y$ axis support of $U(y-y_{m1,2\_n})-U(y-y_{m1,2\_n}-d\_n)$, with $U$ being the step function, indicating the incidence angle $\theta_n$. \\
In the general case $p_{m1}$ and $p_{m2}$ contain both the original detection beam propagating in $\textbf{r}$, and the residual control back wave propagating in $\textbf{r}_b$. Therefore, in order to estimate $A_n$, $\omega_n$, $\theta_n$ and $\phi_{m1,2\_n}$ from $p_{m1}$ and $p_{m2}$, the original source beam first needs to be separated from the total measurement.
In one dimension this is quite a straightforward procedure, common in many wave applications, and can be achieved, for example, using the acoustic transfer matrix method \cite{song2000transfer}. 
In two dimensions it is more complicated, but methods do exist \cite{becker2020real}. 
In our system, however, this separation can occur naturally by two means. The first is geometrical. Since the beam is assumed of finite $y$ axis extension $d$, for a nonzero $\theta$ the original $\psi$ and the backwave become completely geometrically decoupled at $x_d= x_o-\frac{1}{2}d\cot{\theta}$. Therefore, sensor arrays placed to the left of $x_d$ will contain two separate nonzero sections, from which only the lower one needs to be identified. However, as $x_d$ may result too far from the object (due to small $\theta$ and/or large $d$), this type of decoupling is not very reliable. Instead, we assume the second type, a dynamical one, which has the potential to occur due to the non-overlapping time intervals assumption. \\
As illustrated in Fig. \ref{fig:estimation}-Left, full detection wave - backwave separation takes place if the tail of the interval $n$ backwave left the sensor arrays before the head of the interval $n+1$ detection wave entered it. That is, $\alpha_{n+1}\geq \beta_n+2\tau_{mo}$, where $\tau_{mo}=l_{mo}/c$ is the time that takes a wave of speed $c$ to travel $l_{mo}=(x_o-x_{m1})/\cos{\theta}$, the distance between the sensors array at $x_{mo}$ and the actuators array at $x_o$ in the $\textbf{r}$ direction. 
Since the sensors location is our choice, $\tau_{mo}$ can be made reasonably small. Then, the parameters estimation needs to take place during the time when the interval $n$ detection wave head entered the sensor arrays and before the backwave of the same interval reaches these arrays, i.e. during $\alpha_n\leq t \leq \alpha_n+2\tau_{mo}$, as illustrated in Fig. \ref{fig:estimation}-Right. We denote this time period by the effective measurement time. 
The consequent estimation of $A_n$, $\omega_n$, $\theta_n$ and $\phi_{m1,2\_n}$ from the separated source beam can be carried out using any appropriate signal processing algorithm, and we assume here that these parameters are available and fed into $C_\psi$. \\
At each time step $n$ the controller $C_\psi$ then predicts the expected pressure fields at the actuators arrays. 
In particular, it predicts which actuators along the arrays at $x_{o,c}$ and $x_{o,c}-\epsilon$ need to be activated, and generates the corresponding support windows $U_{o,c\_n}(y)$, as well as a Gaussian envelope $G_n(y)$ of scaling $\sigma$ that smooths out the windows corners, given by
\begin{equation}  \label{eq:step}
    U_{o,c\_n}(y)=U(y-y_{o,c\_n})-U(y-(y_{o,c\_n}+d\_n))  \quad , \quad G(y)=e^{-\frac{1}{2}\left(\sigma y\right)^2}.
\end{equation}
Here, $y_{o,c\_n}$ refers to $y_{o1,2\_n}$ for the dead zone opening part and to $y_{c1,2\_n}$ for its closing.
The controller $C_\psi$ also needs to predict the phase shift between the measurement position and the control position. This is obtained via
\begin{equation}
    \phi_{o,c\_n}=\frac{x_{o,c}-x_{m1,2}}{c \cos{\theta_n}}\omega_n.
\end{equation}
The next stage is determining the controllers $C_{fo}$ and $C_{fc}$.
The control waves $f_{o,c}(y,t)$ need to account for the dynamics introduced by the function $Q(s)$ defined in \eqref{eq:Q}, which is responsible for the  unidirectional propagation, and stems from the control law $C_h$, defined in \eqref{eq:C_h}. 
The amplitudes of $f_{o,c}(y,t)$ and their phases are thus respectively modulated by $M_{Q\_n}$, the inverse of the amplitude of $Q(s)$, and $\delta\phi_{Q\_n}$, the negative phase of $Q(s)$. Both are calculated at the detection frequency $\omega_n$, and are given by
\begin{equation}
    M_{Q\_n}=\frac{1}{|Q(i\omega_n)|} \quad , \quad \delta\phi_{Q\_n}=-\sphericalangle Q(i\omega_n).
\end{equation}
The controllers $C_{fo}$ and $C_{fc}$ then need to determine the control wave rotation at the angle $\theta_n$. This can be obtained using a phased array like modulation of the form 
\begin{equation} \label{eq:varphi}
    \Delta\varphi_n(y)=\frac{y\omega_n}{c}\sin{\theta_n}.
\end{equation}
Combining \eqref{eq:step}-\eqref{eq:varphi}, the control commands of $C_{fo}$ and $C_{fc}$ become
\begin{equation}  \label{eq:f_final}
    f_{o,c\_n}(y,t)=\sum_n{U_{o,c\_n}(y)G_{o,c\_n}(y)M_{Q\_n}A_ne^{i\omega_n t}e^{i\left(\Delta\varphi_n(y)-\delta \phi_{o,c\_n}-\delta\phi_{Q\_n}\right)}},
\end{equation}
which completes the derivation of the control algorithm.

\section{Cloak performance demonstration}   \label{Time_sim}

\begin{figure*}[htbp] 
\begin{center}
\setlength{\tabcolsep}{0pt}
  \begin{tabular}{l l}
 \textbf{(a)} & \textbf{(f)} \\ \includegraphics[height=6.2cm, valign=c]{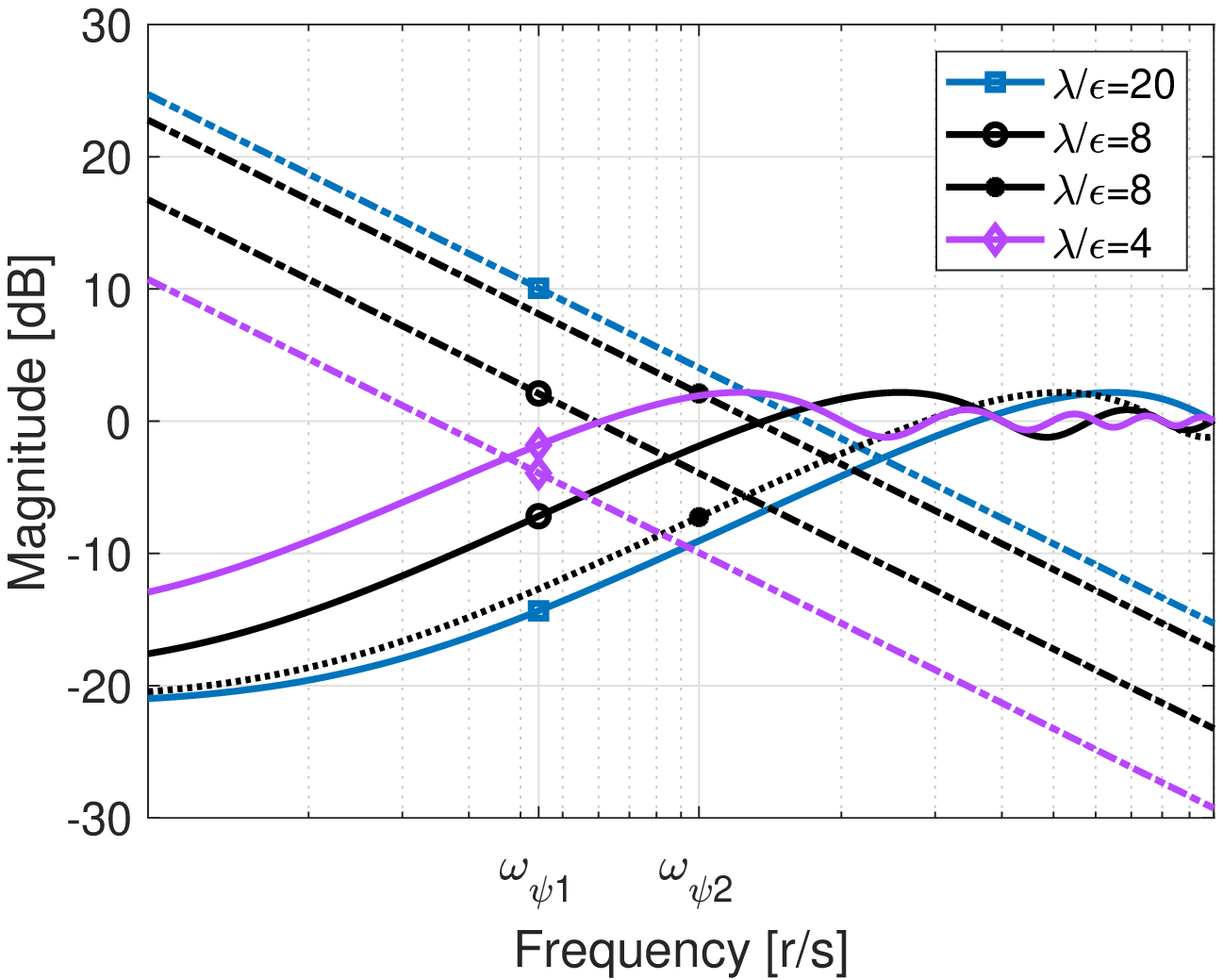}   & \includegraphics[height=6.2 cm, valign=c]{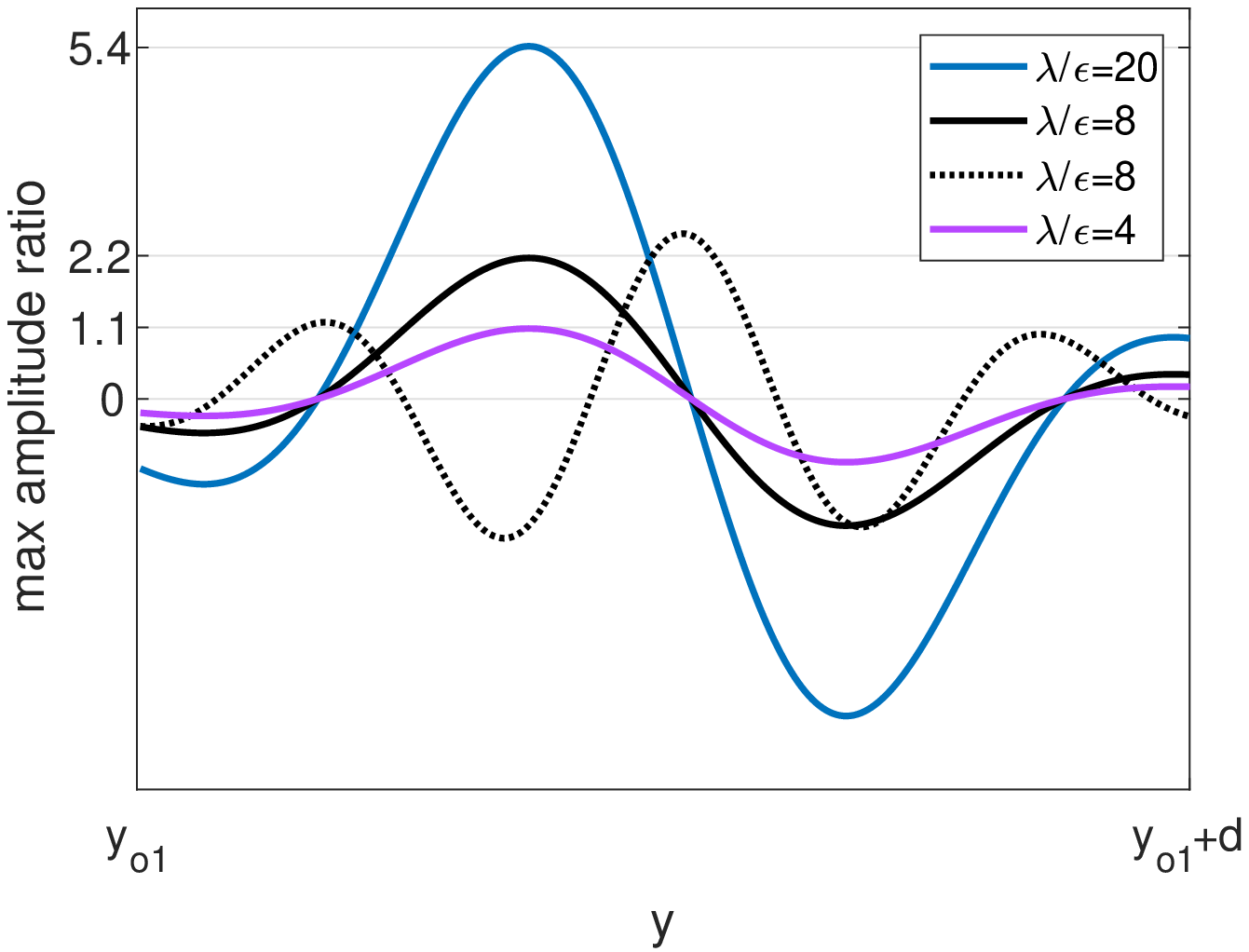}
 \end{tabular}
 \begin{tabular}{c c c c}
 \textbf{(b)} $\lambda/\epsilon=20$ & \textbf{(c)} $\lambda/\epsilon=8$ & \textbf{(d)} $\lambda/\epsilon=8$ & \textbf{(e)} $\lambda/\epsilon=4$ \\ 
 \includegraphics[height=4.2 cm, valign=c]{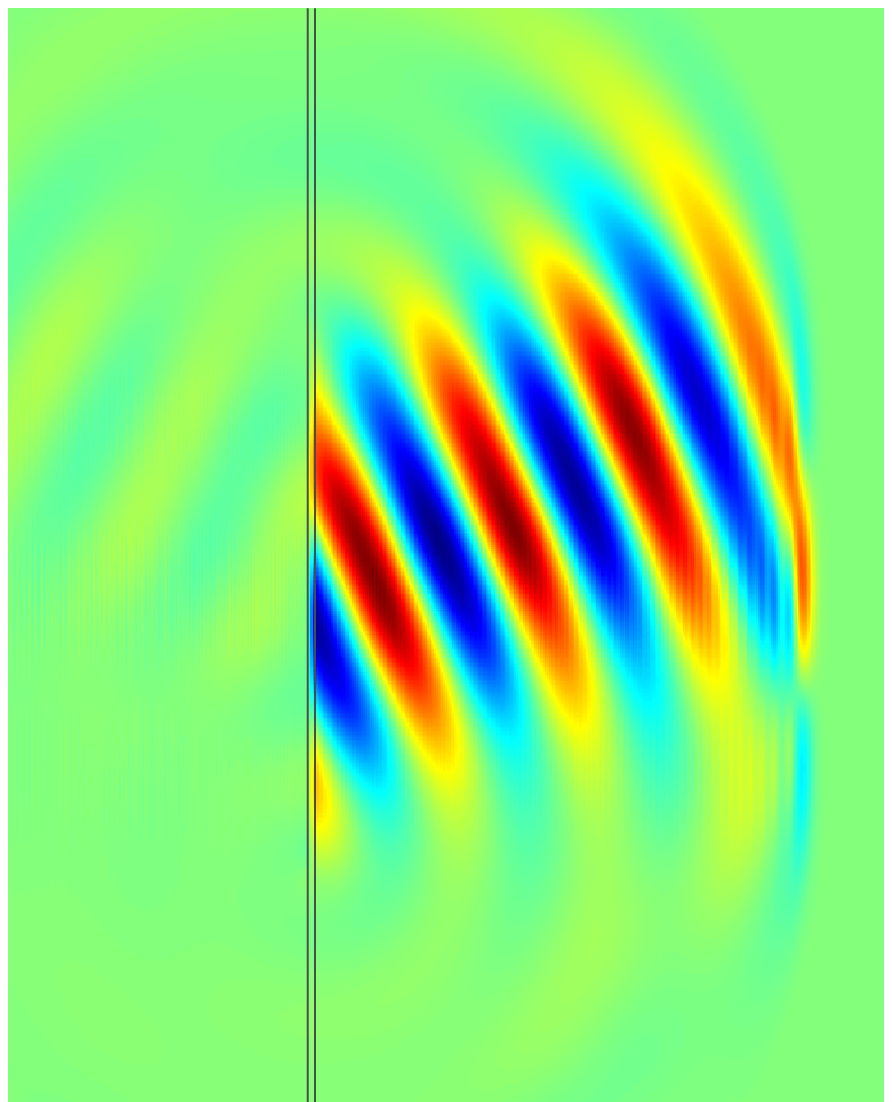} &
  \includegraphics[height=4.2 cm, valign=c]{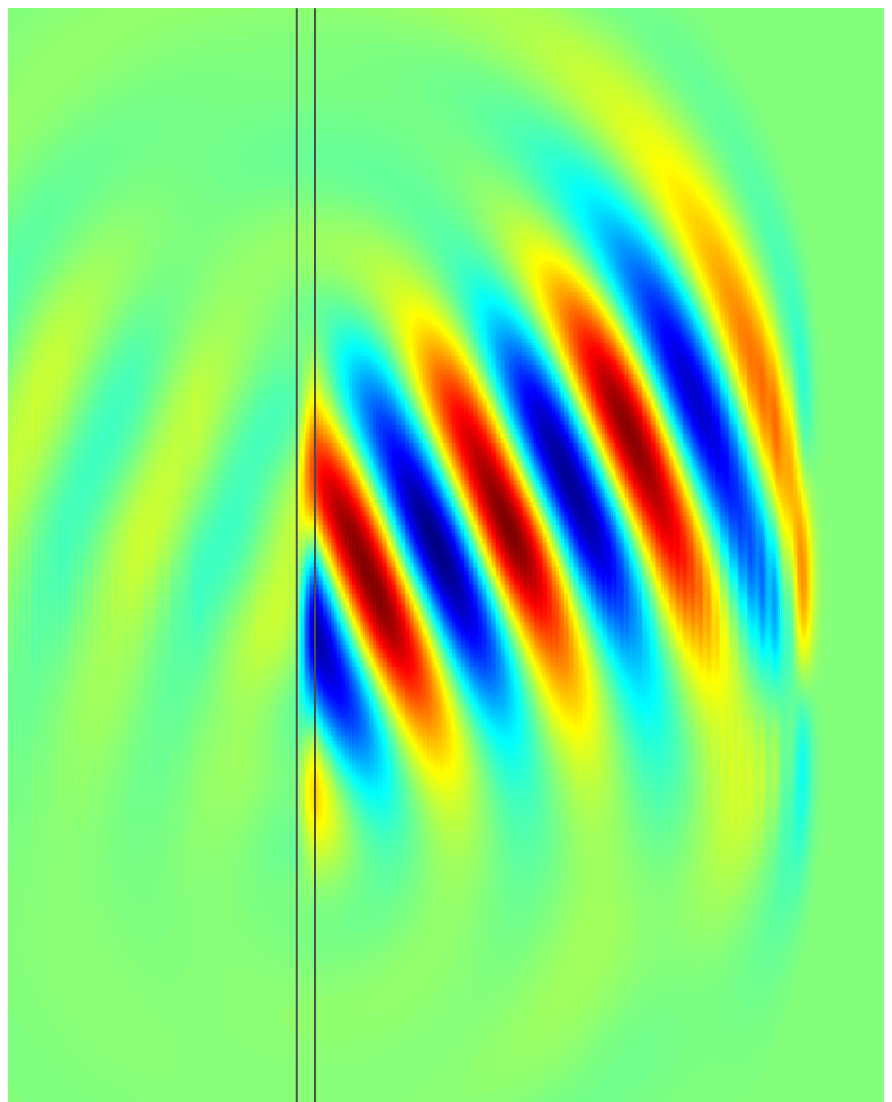}  &  \includegraphics[height=4.2 cm, valign=c]{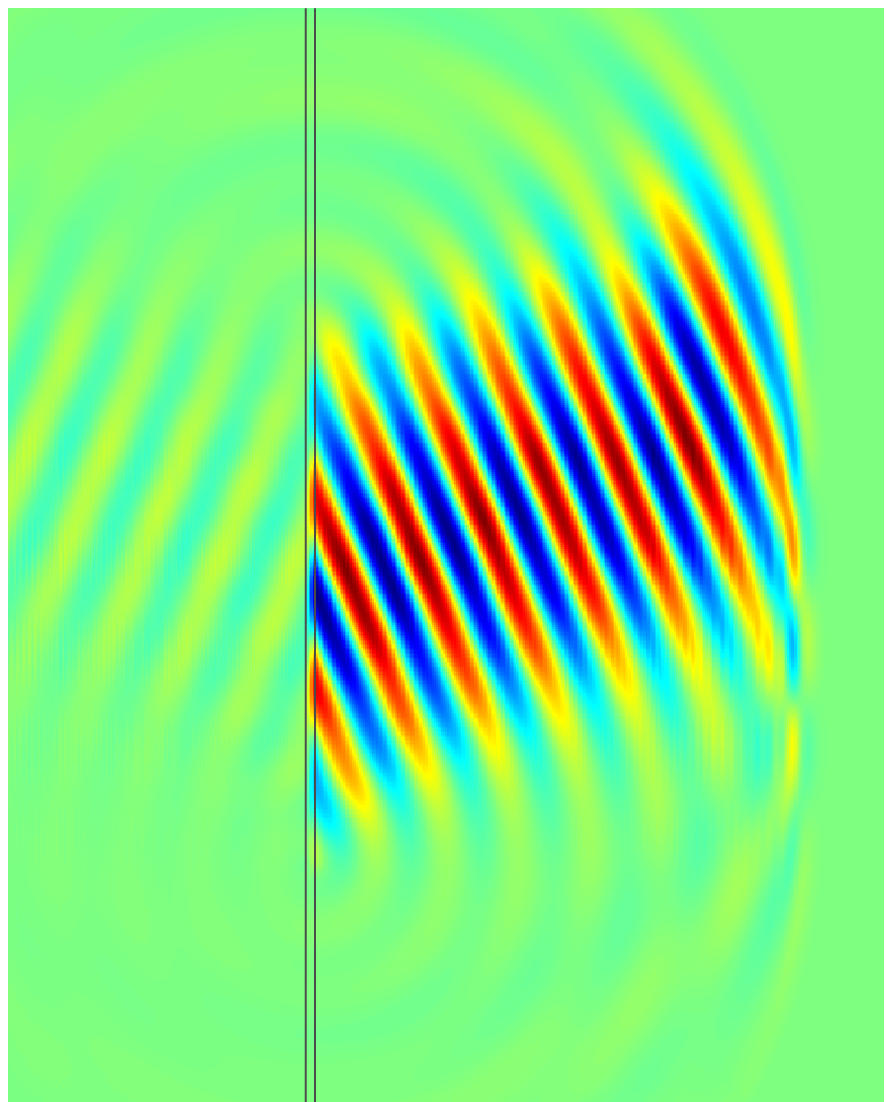} &
     \includegraphics[height=4.2 cm, valign=c]{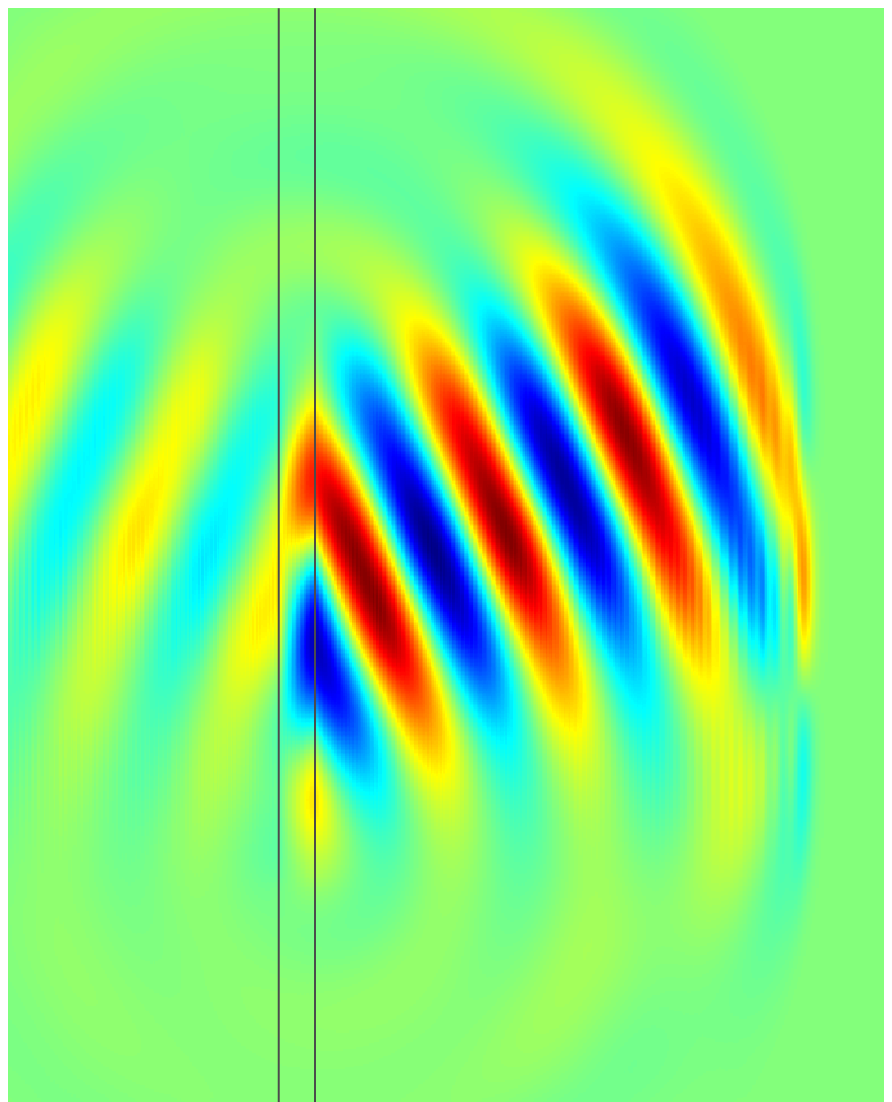} \\
     & & &
\end{tabular} 
\includegraphics[width=4.0 cm]{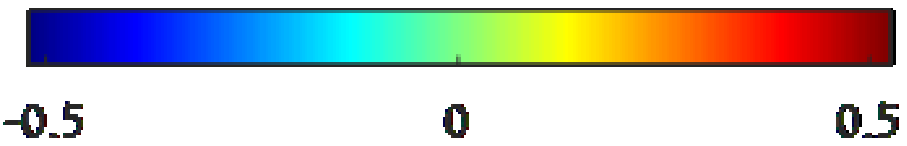}
\end{center}
 \caption{Trade-off between the control effort and the control wave uni-directionality level. (a) Frequency response diagram of the controller $C_h(s)$, dashed-dotted, and of the backwave $\overline{Q}(s)$, solid, for $\epsilon=0.02$ m, blue, $\epsilon=0.05$ m, solid-black, $\epsilon=0.025$ m, dotted-black, and $\epsilon=0.02$ m, purple. The markers on the curves indicate the amplitudes corresponding to the source frequencies $\omega_{\psi1}=0.86$ kHz and $\omega_{\psi2}=1.72$ kHz. (b)-(e), Time domain snapshots of control signal $h(y,t)$ according to the algorithm in \eqref{eq:C_h} for $\theta=22^o$. (b),(c),(e), $\lambda=0.4$ m with $\epsilon=0.02,0.05,0.1$ m, leading to $\lambda/\epsilon=20,8,4$. (d), $\lambda=0.2$ m with $\epsilon=0.025$ m, leading to $\lambda/\epsilon=8$. (f), $h_o(y,t)$ along the $y$ axis actuation section at a time instance when the amplitude of $h_o(y,t)$ is maximal. The higher is $\lambda/\epsilon$, the smaller is the backwave, but the higher is the control effort.}
\label{trade_off}
\end{figure*}

\begin{figure*}[htbp] 
\begin{center}
\setlength{\tabcolsep}{1pt}
 \begin{tabular}{c c c c c}
 $\frac{\lambda}{\epsilon}$ & & Free space & Cloak ON & Observers \\
 4 & & \includegraphics[height=4.2 cm, valign=c]{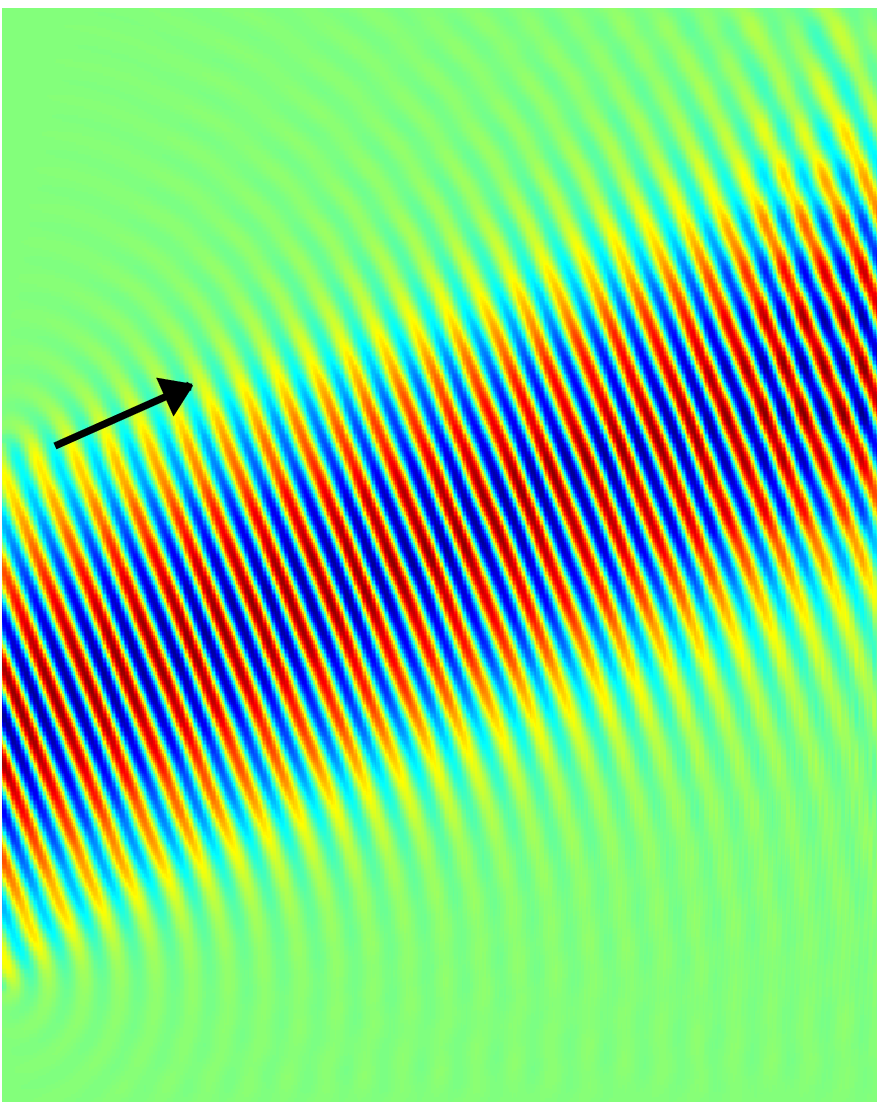} &
  \includegraphics[height=4.2 cm, valign=c]{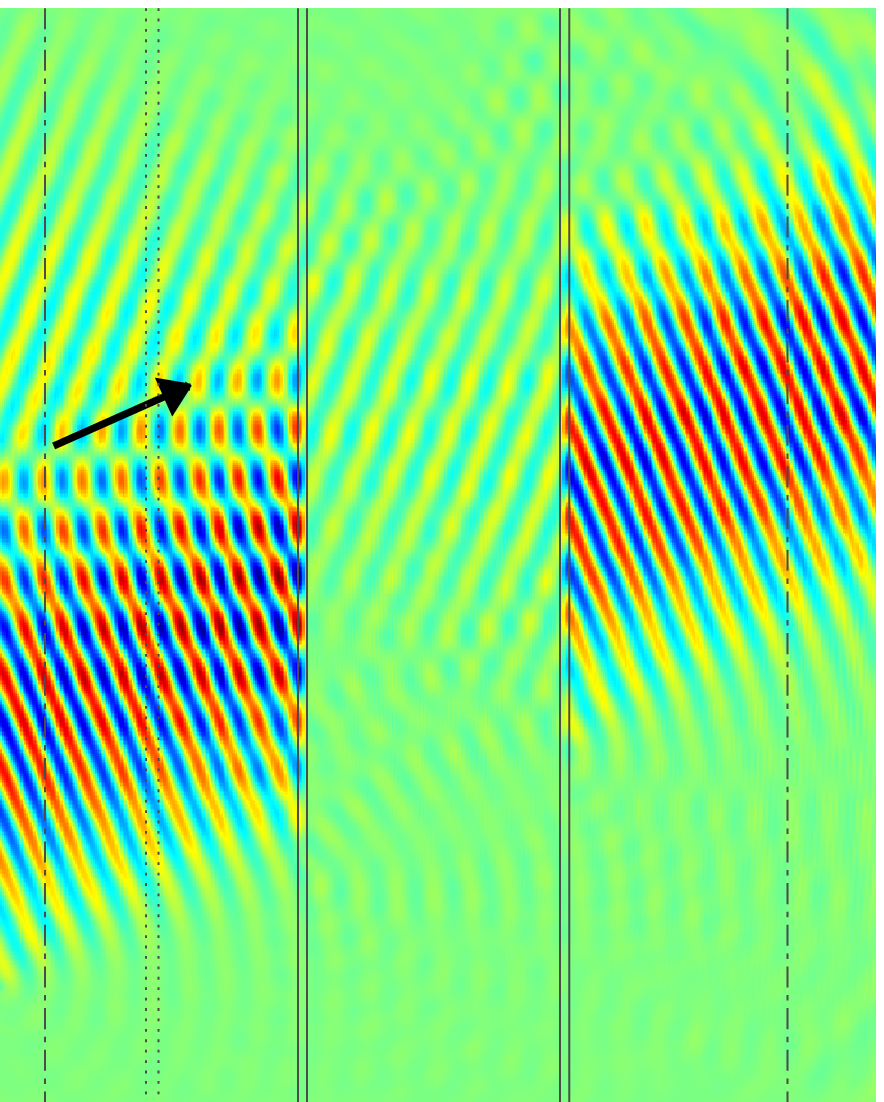}  &  \includegraphics[height=4.9 cm, valign=c]{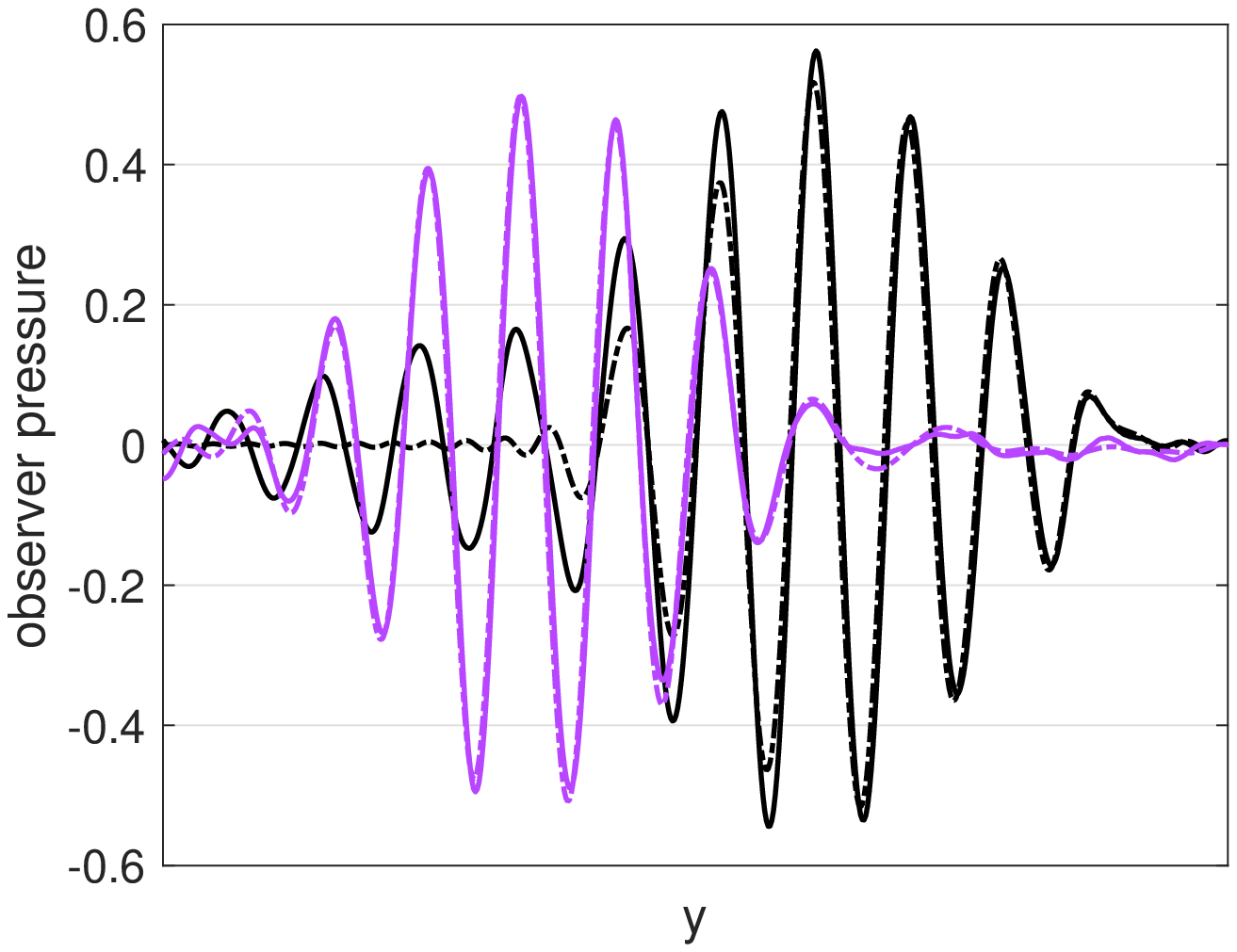} \\

  8 & & \includegraphics[height=4.2 cm, valign=c]{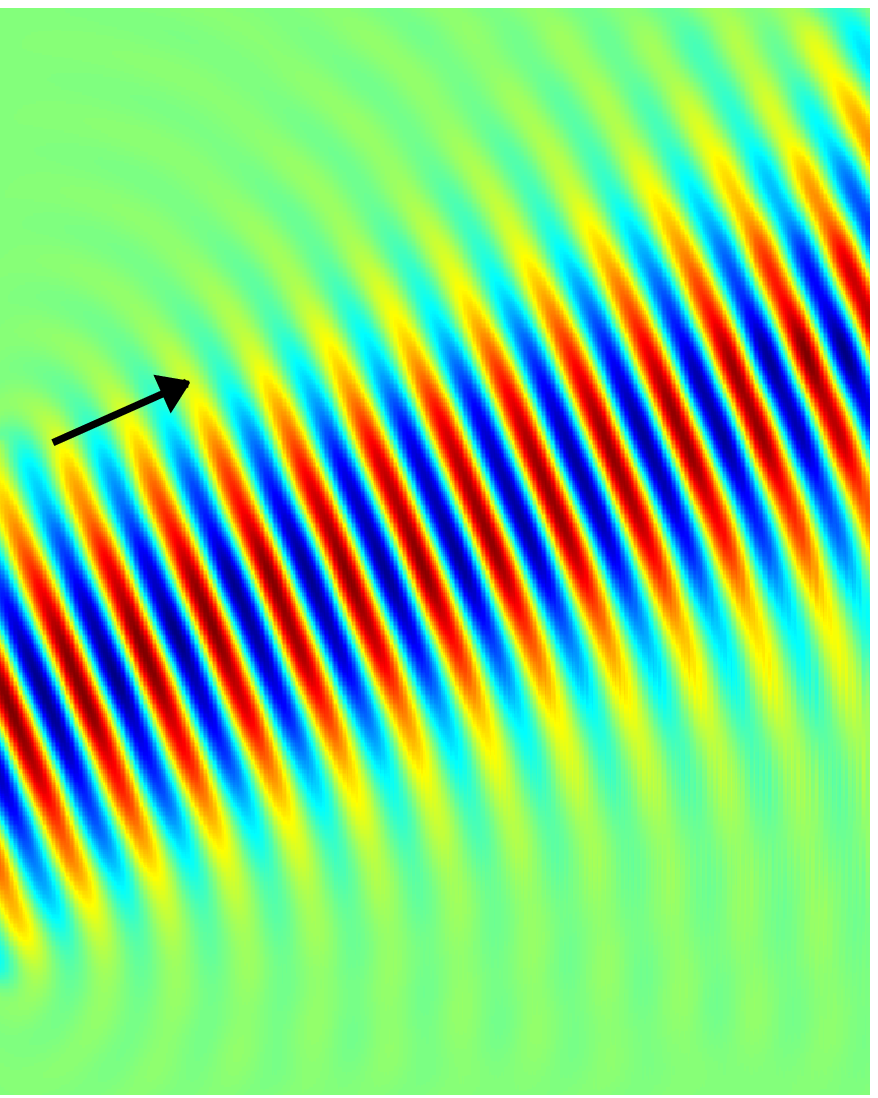} &
  \includegraphics[height=4.2 cm, valign=c]{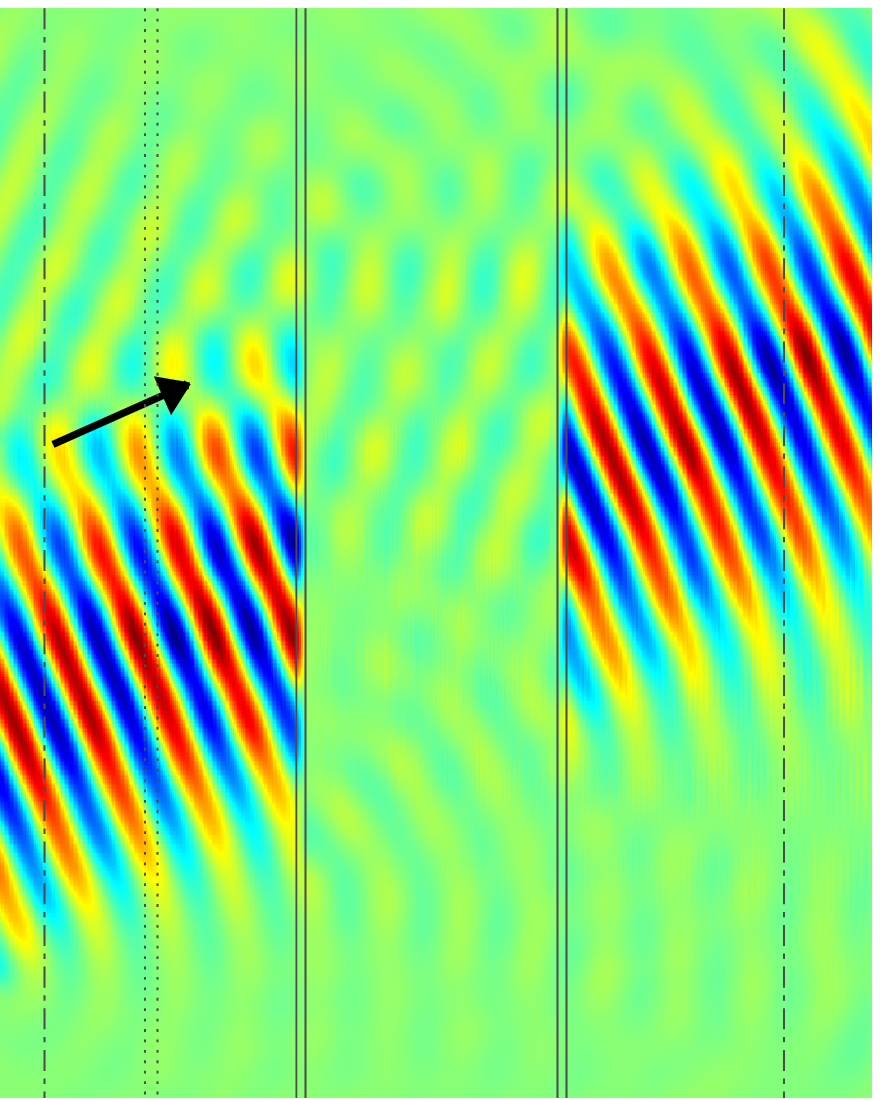}  &  \includegraphics[height=4.9 cm, valign=c]{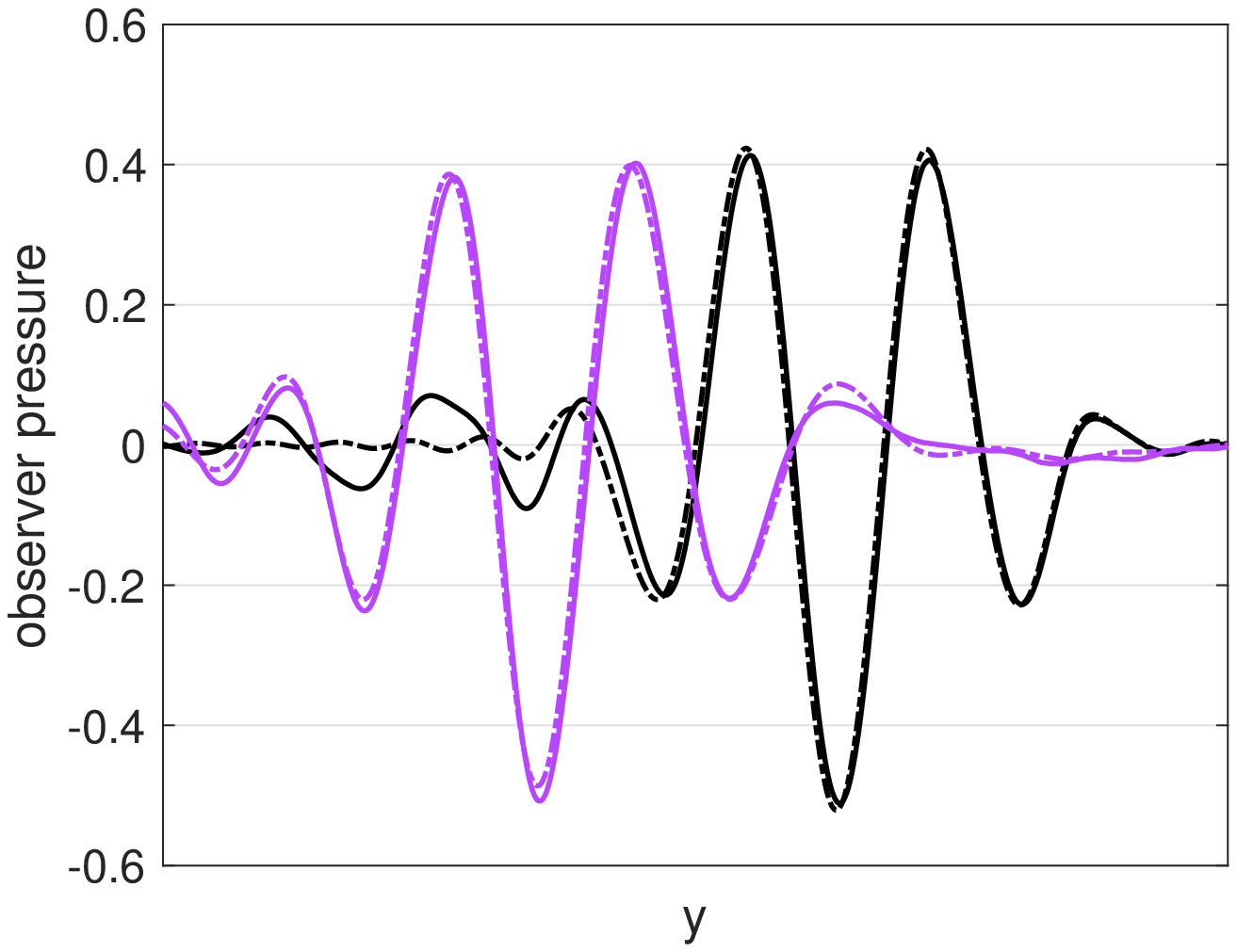} \\
 
  12 & & \includegraphics[height=4.2 cm, valign=c]{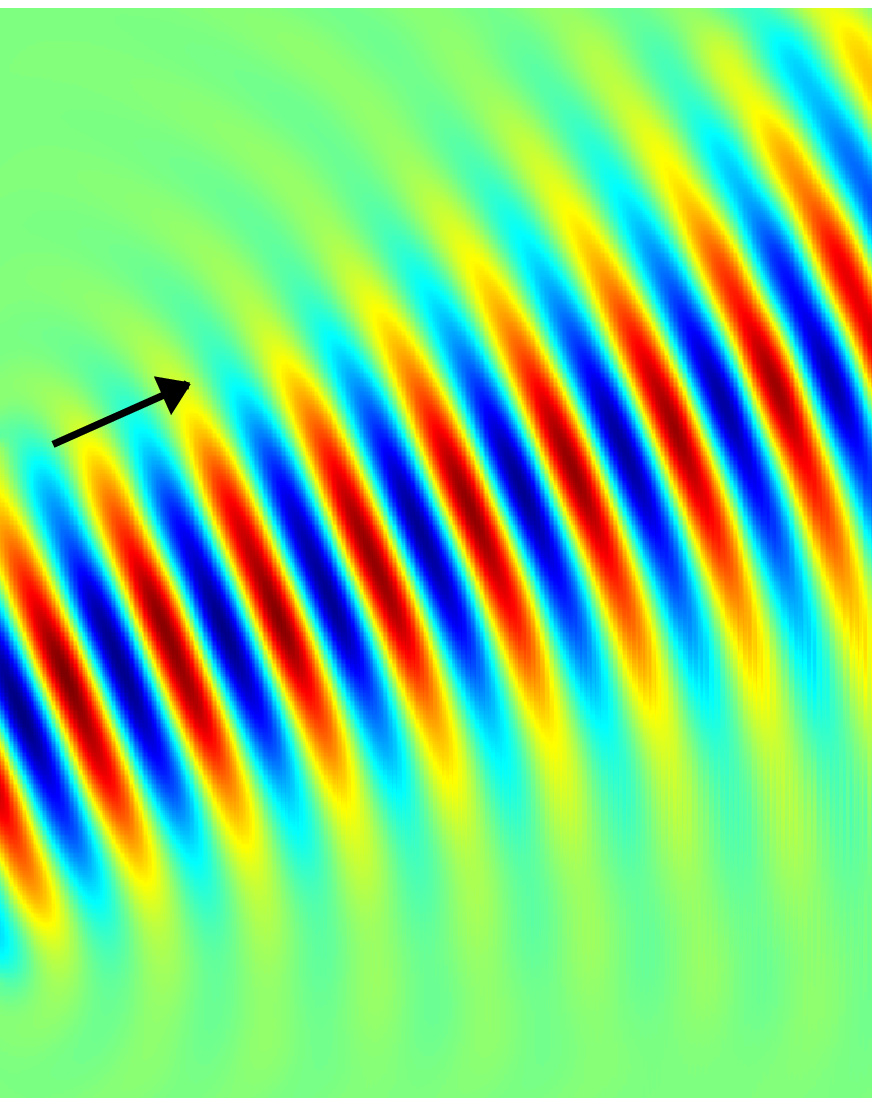} &
  \includegraphics[height=4.2 cm, valign=c]{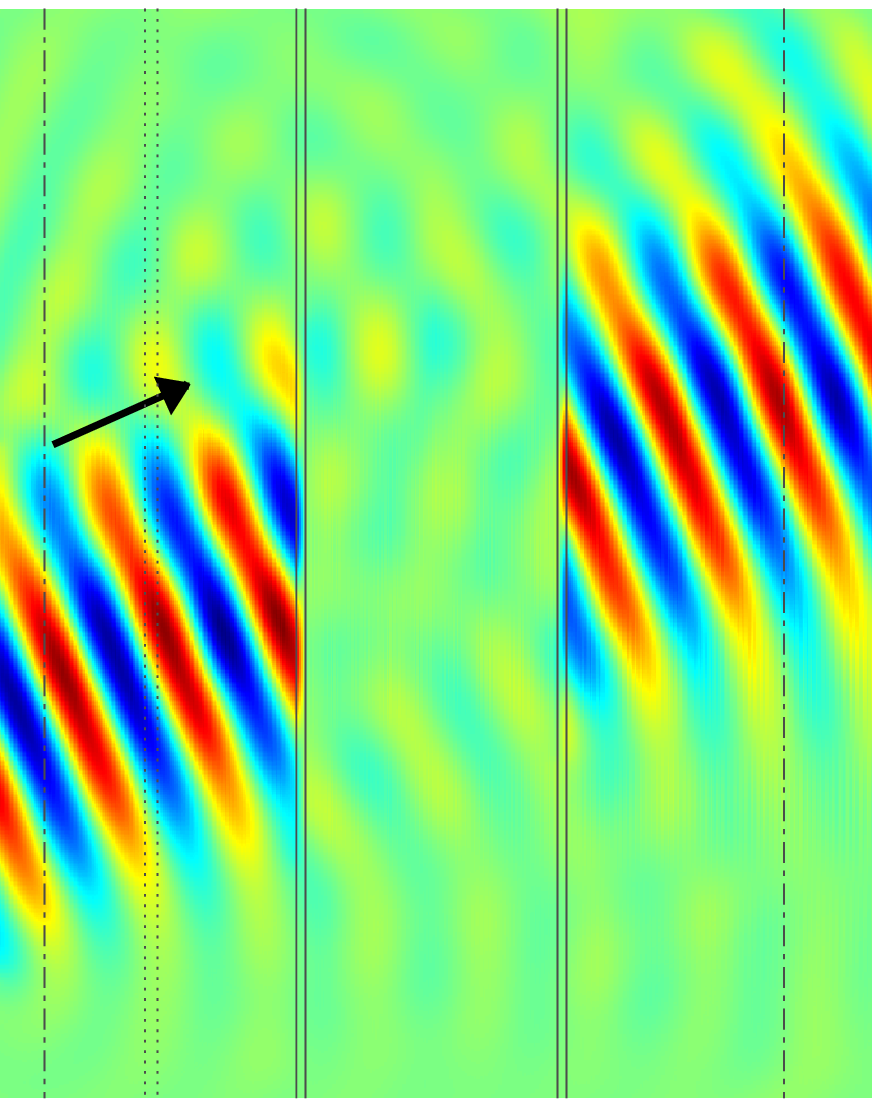}  &  \includegraphics[height=4.9 cm, valign=c]{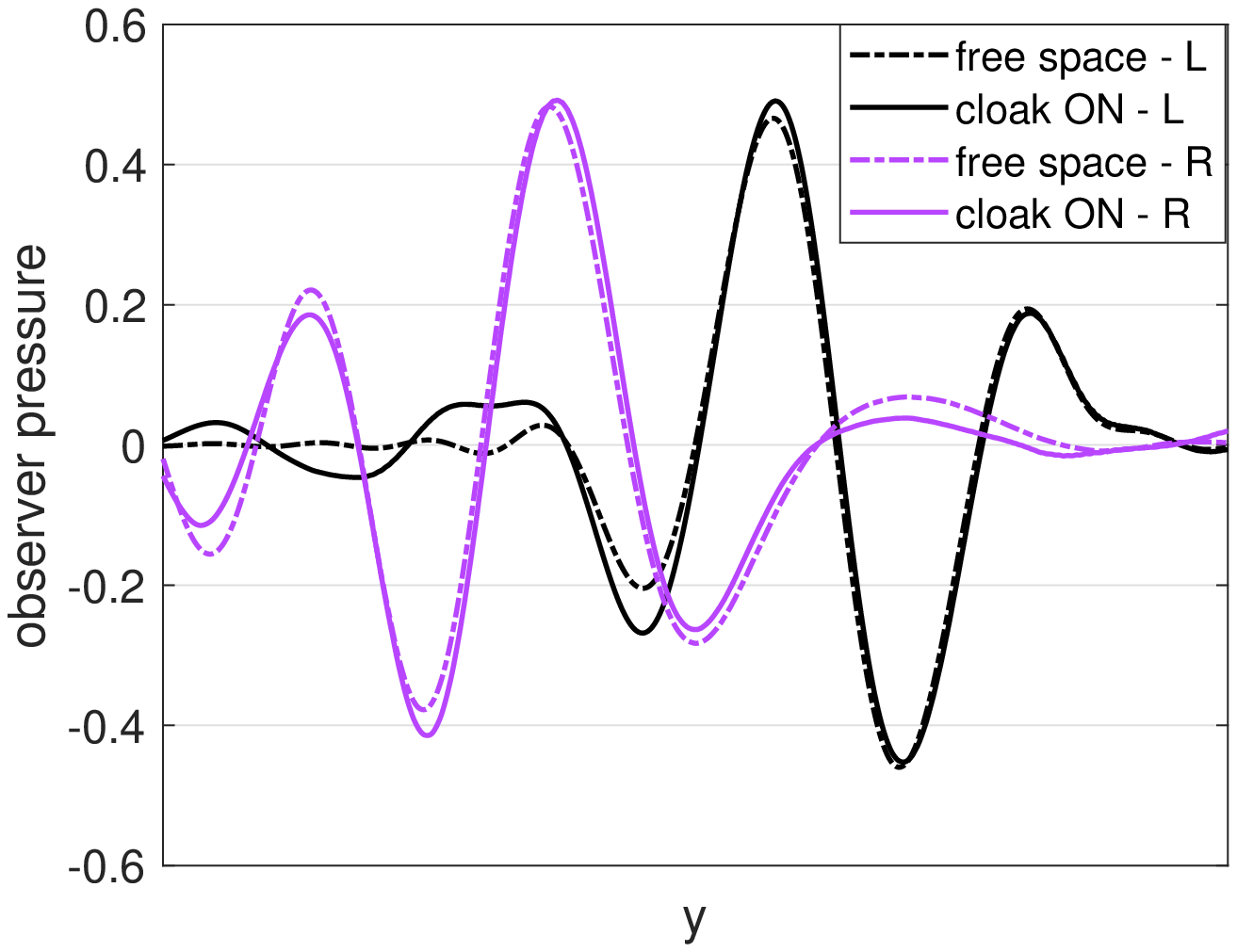} \\
 
  16 & & \includegraphics[height=4.2 cm, valign=c]{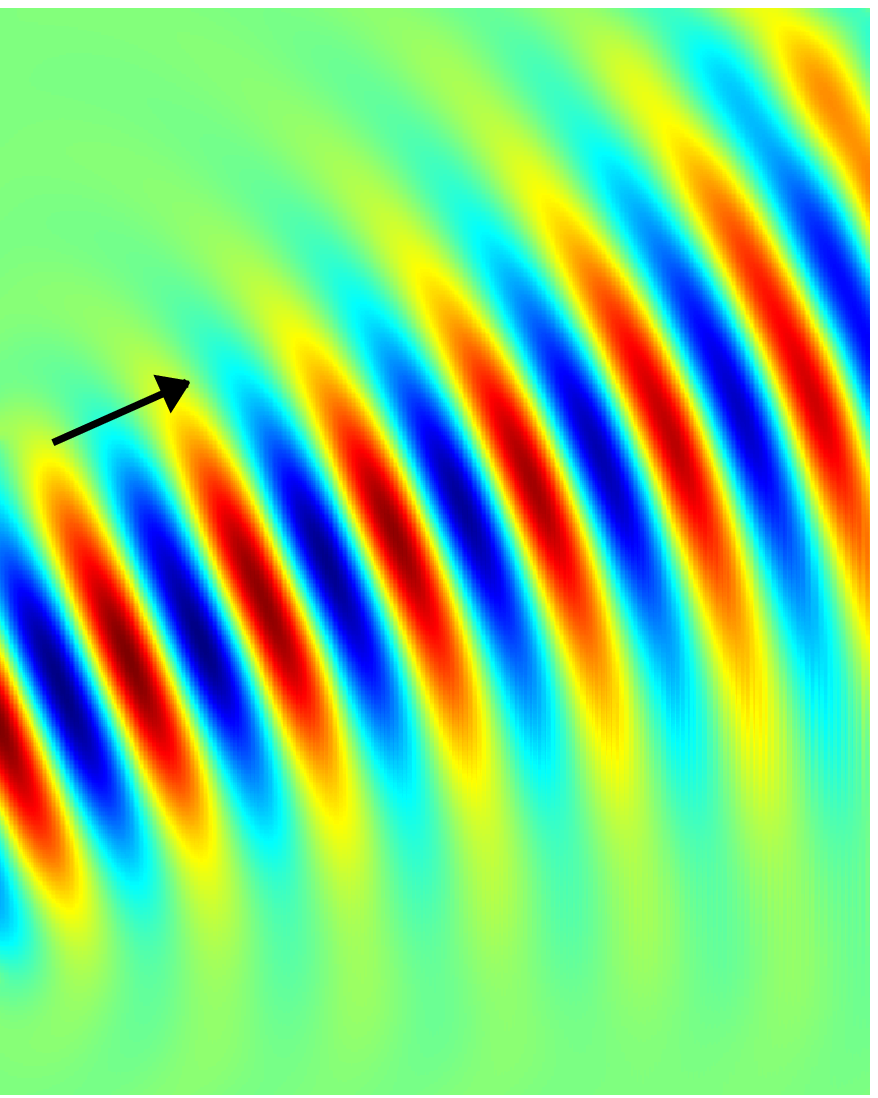} &
  \includegraphics[height=4.2 cm, valign=c]{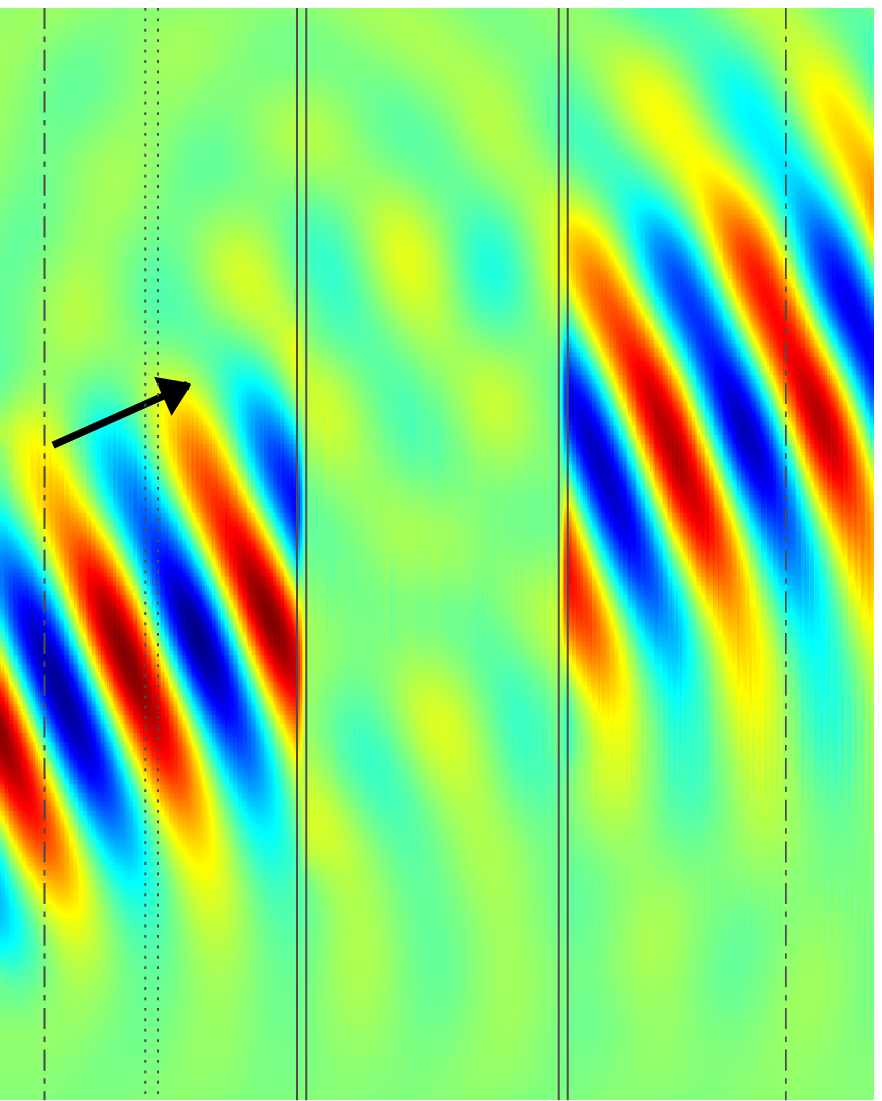}  &  \includegraphics[height=4.9 cm, valign=c]{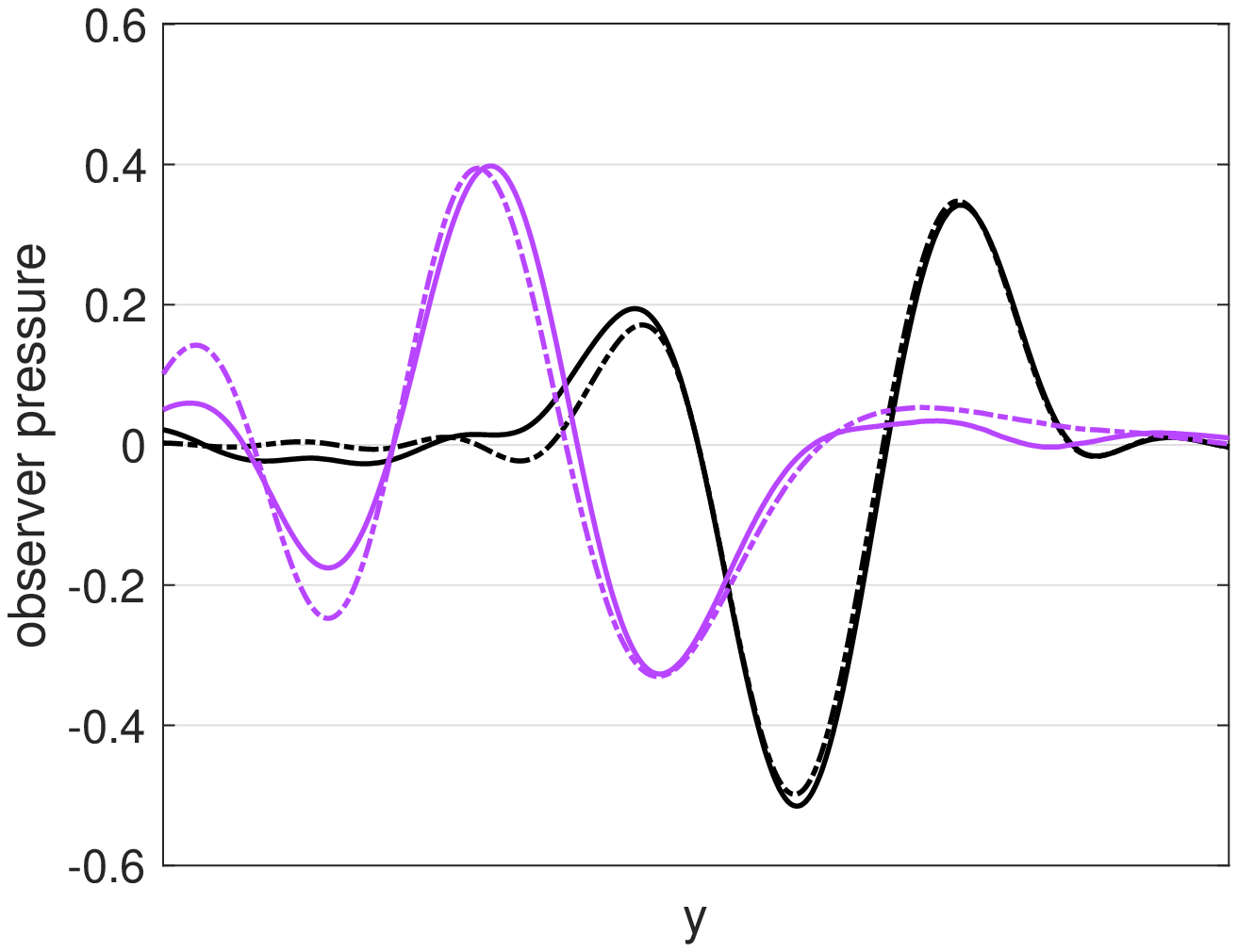} 
\end{tabular} 
\includegraphics[width=4.0 cm]{Figures/colorbar_jet.eps}
\end{center}
 \caption{Cloak performance demonstration for a varying detection frequency $\omega$. Plotted are snapshots of the pressure field time responses to a detection beam emitted from the left boundary of a $2.4\times 3.0$ m$^2$ air waveguide with $\epsilon=0.025$ m. Rows 1:4 correspond to source frequencies $\omega=3.4, 1.72, 1.14, 0.86$ kHz implying the ratio $\lambda/\epsilon=4,8,12,16$. Left column: entire waveguide responses when the cloak is off, giving free space propagation (black arrow - propagation direction). Middle column: entire waveguide responses when the cloak is on, resulting in the required dead zone formation. Vertical lines: solid - the actuators, dotted - the sensors, dashed-dotted - the observers. Right column: comparison of cloak-on (solid) and cloak-off (dashed-dotted) responses at the left (black) and right (purple) observers.}
\label{lambda_change}
\end{figure*}

\begin{figure*}[htbp] 
\begin{center}
\setlength{\tabcolsep}{1pt}
 \begin{tabular}{c c c c c}
 $\theta$ & & Free space & Cloak ON & Observers \\
 $30^\textrm{o}$ & & \includegraphics[height=4.2 cm, valign=c]{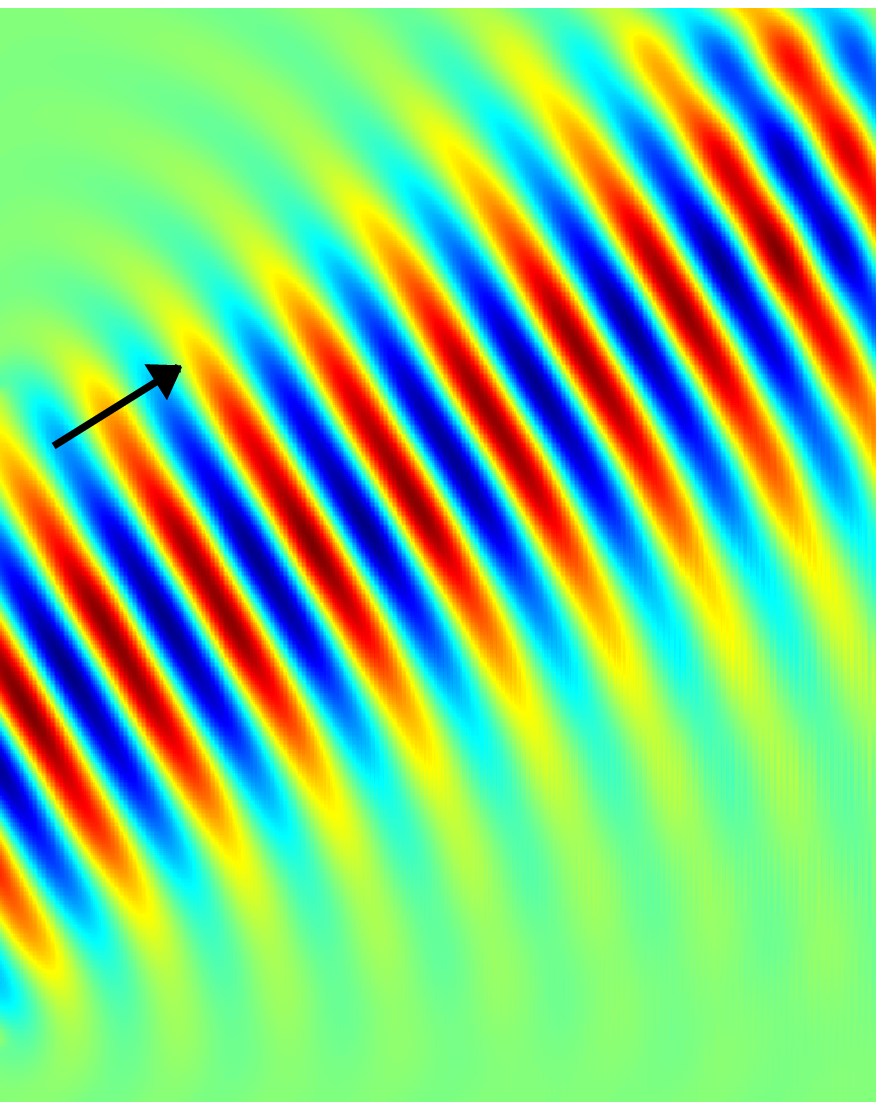} &
  \includegraphics[height=4.2 cm, valign=c]{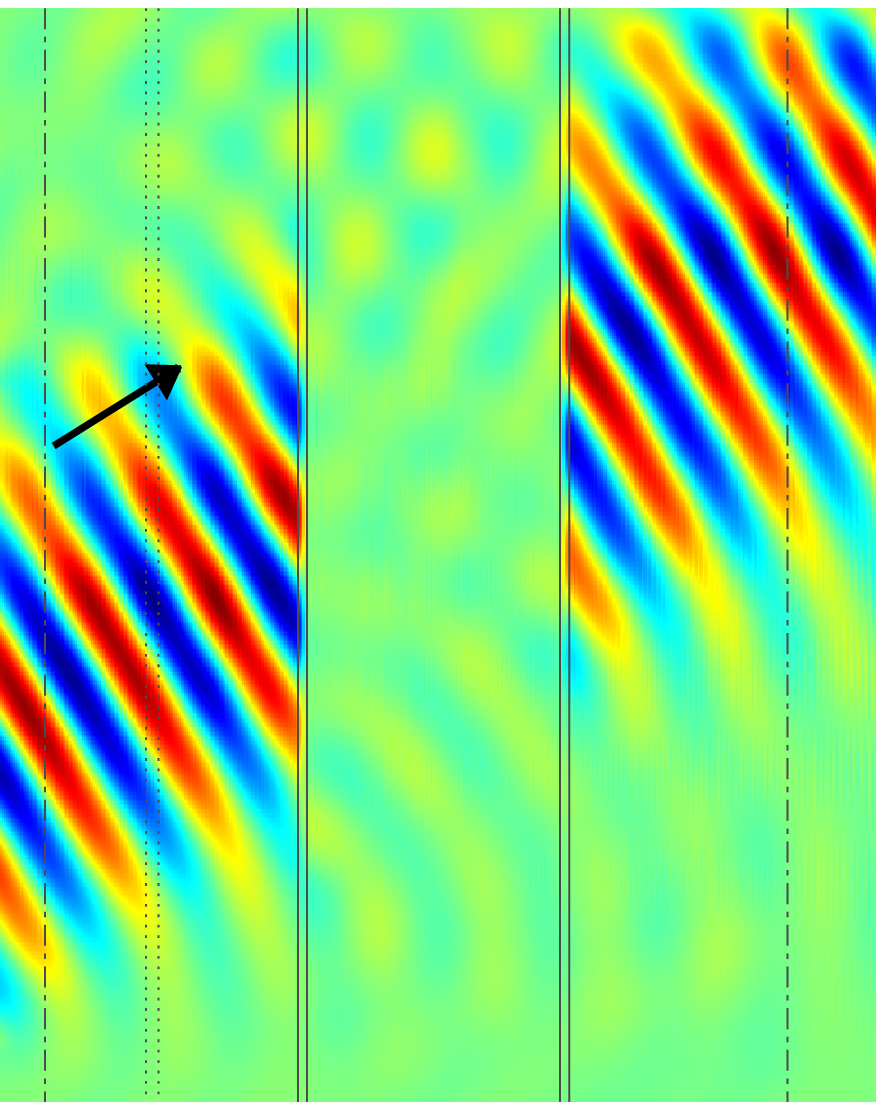}  &  \includegraphics[height=4.9 cm, valign=c]{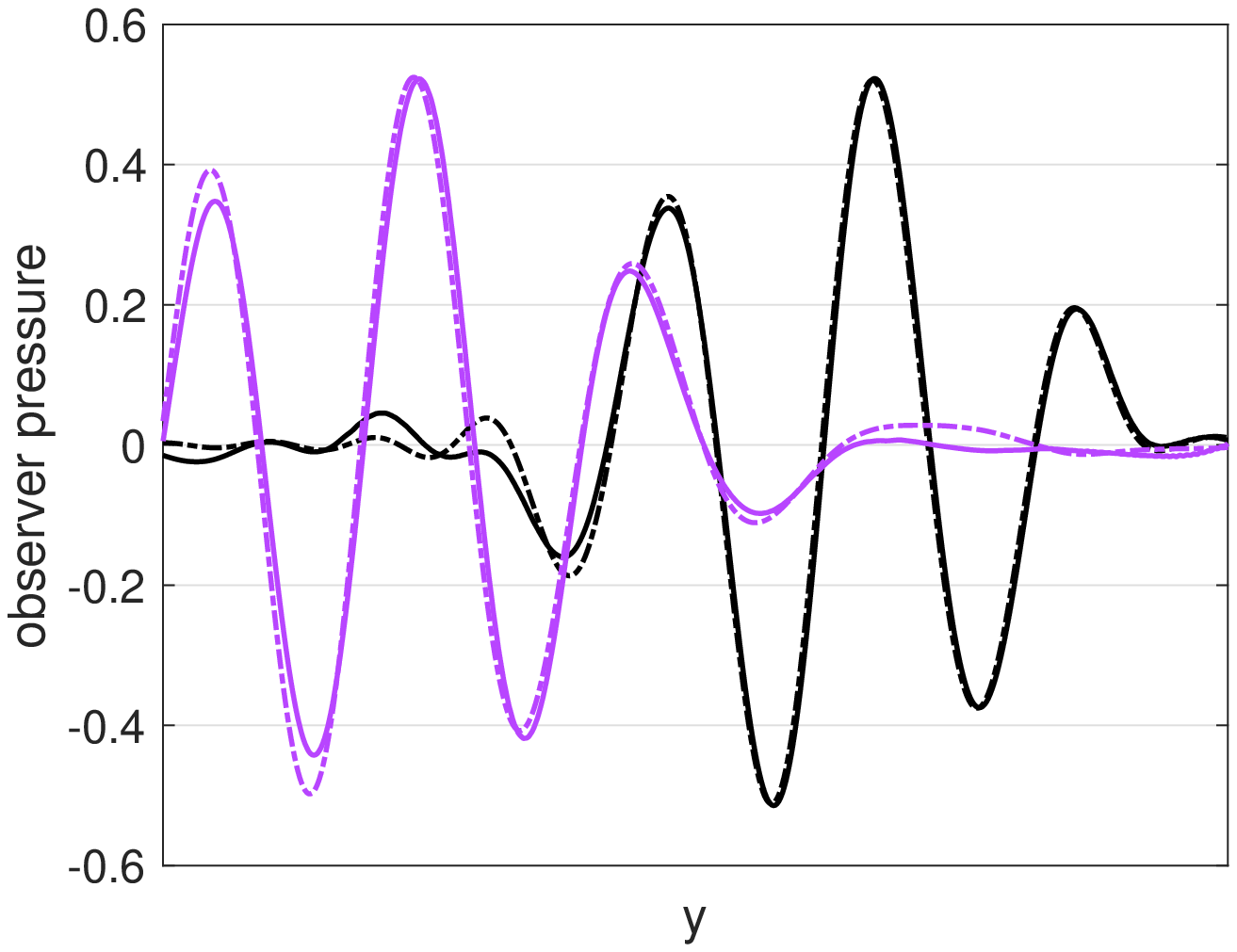} \\

  $15^\textrm{o}$ & & \includegraphics[height=4.2 cm, valign=c]{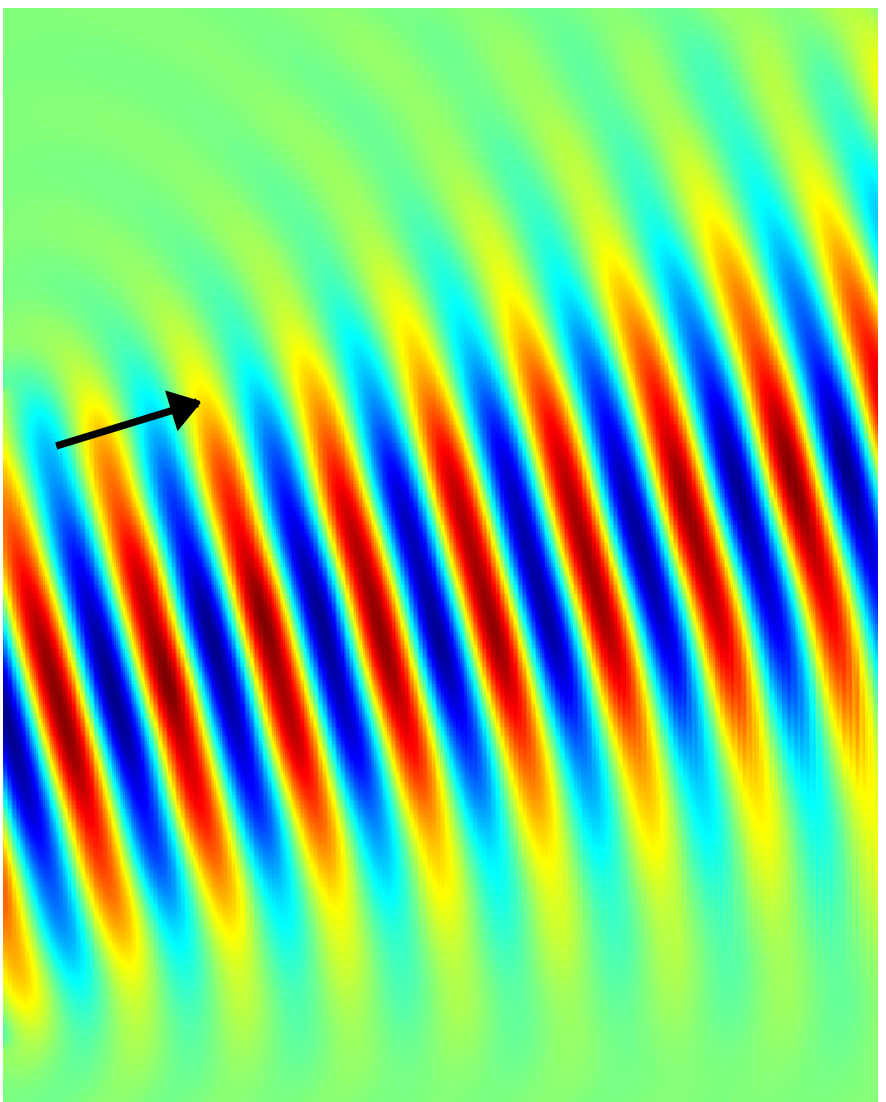} &
  \includegraphics[height=4.2 cm, valign=c]{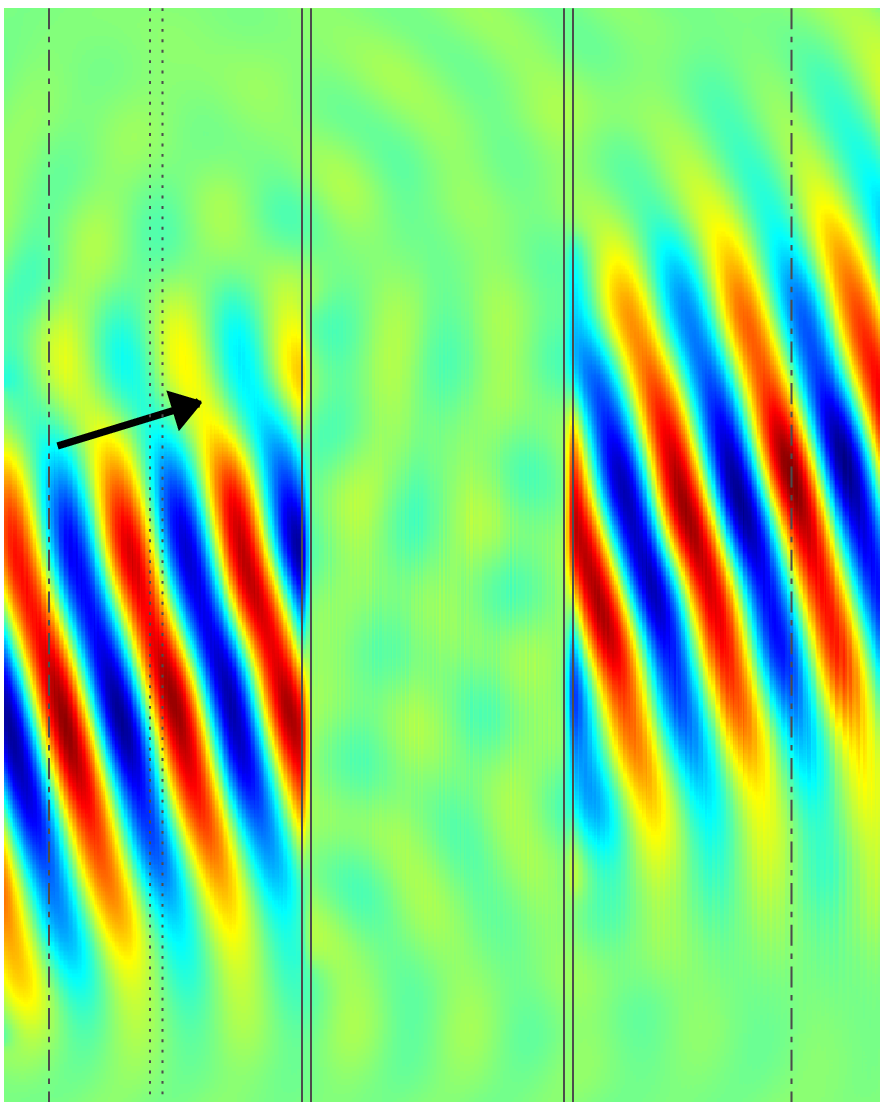}  &  \includegraphics[height=4.9 cm, valign=c]{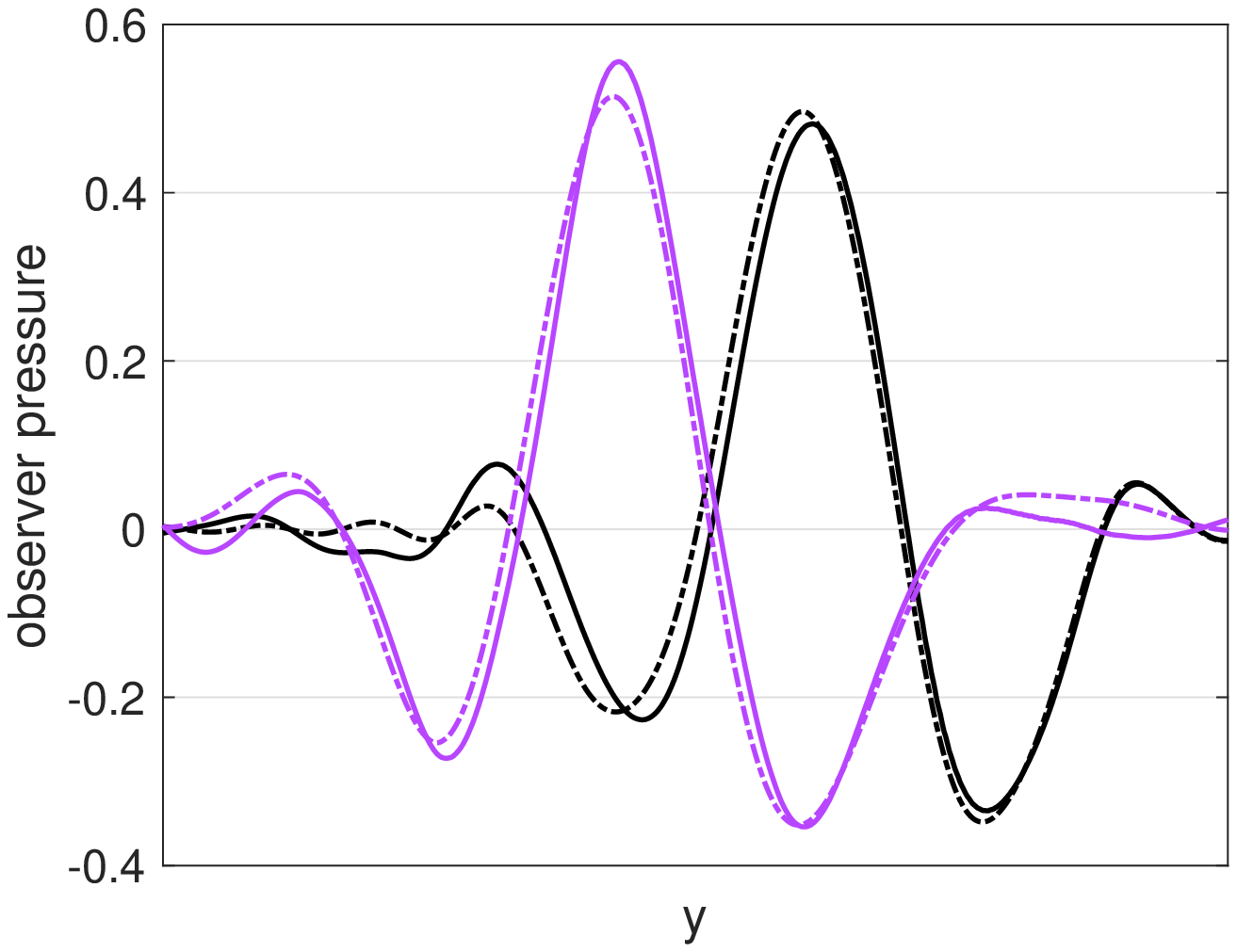} \\
 
  $0^\textrm{o}$ & & \includegraphics[height=4.2 cm, valign=c]{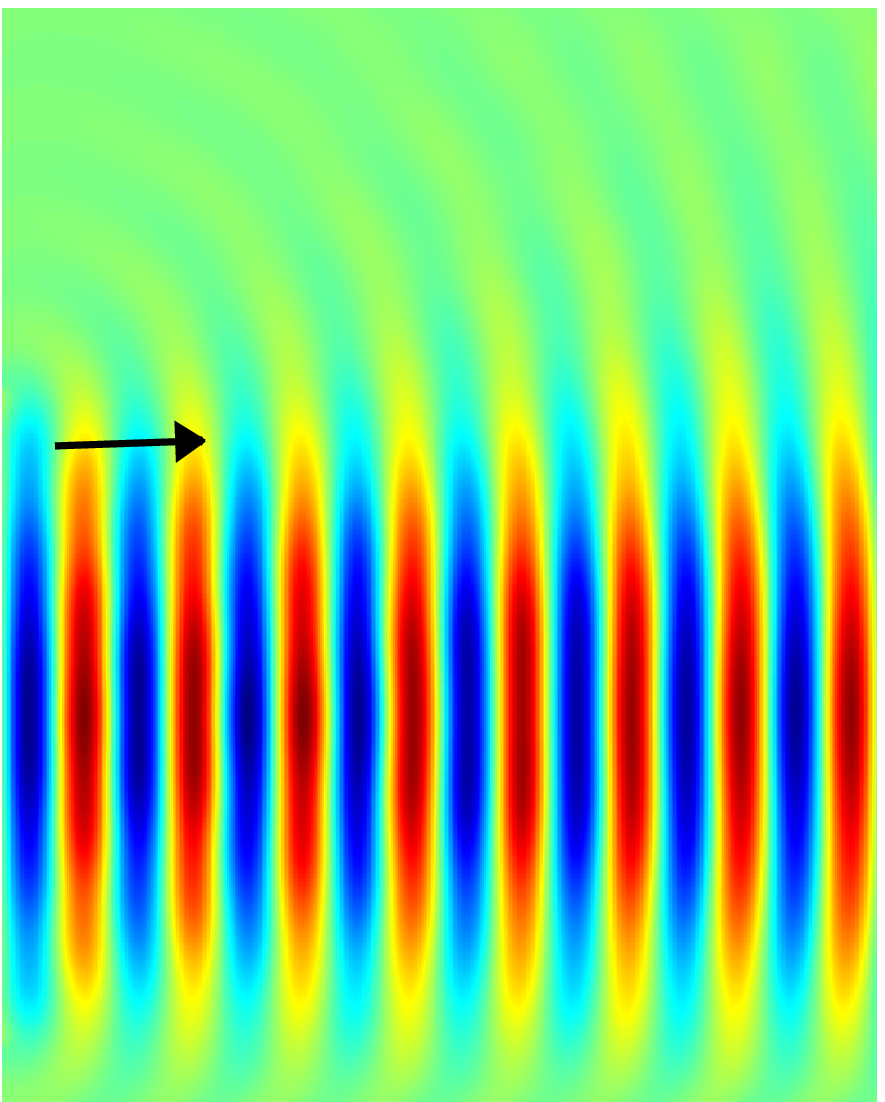} &
  \includegraphics[height=4.2 cm, valign=c]{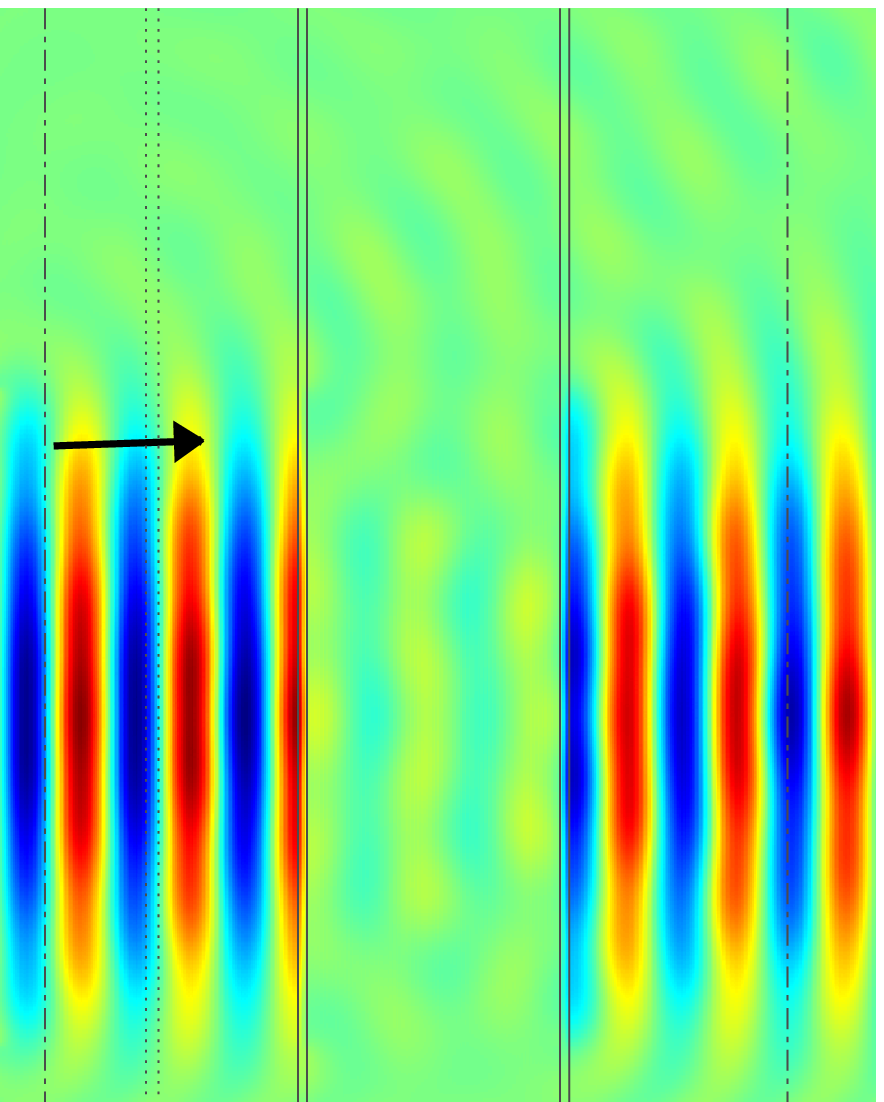}  &  \includegraphics[height=4.9 cm, valign=c]{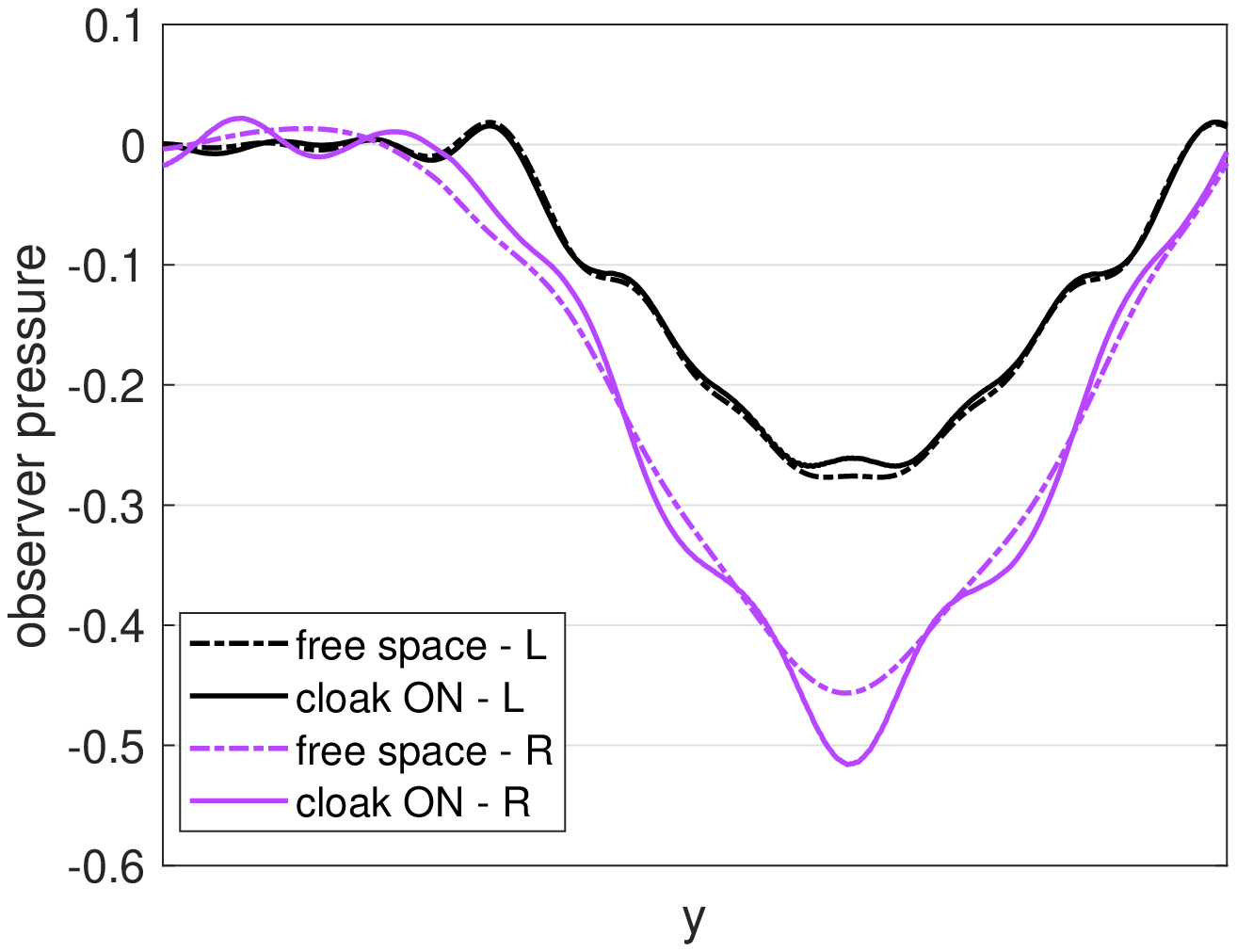} \\
 
  $-30^\textrm{o}$ & & \includegraphics[height=4.2 cm, valign=c]{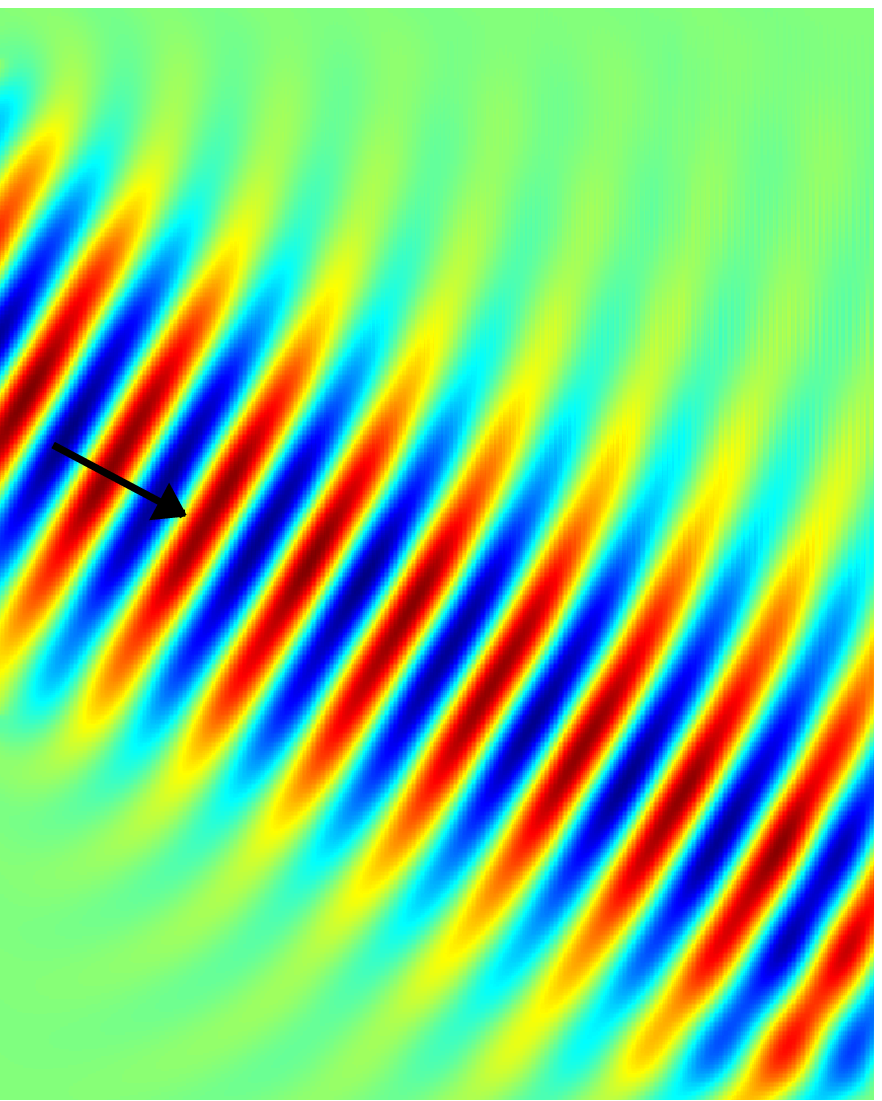} &
  \includegraphics[height=4.2 cm, valign=c]{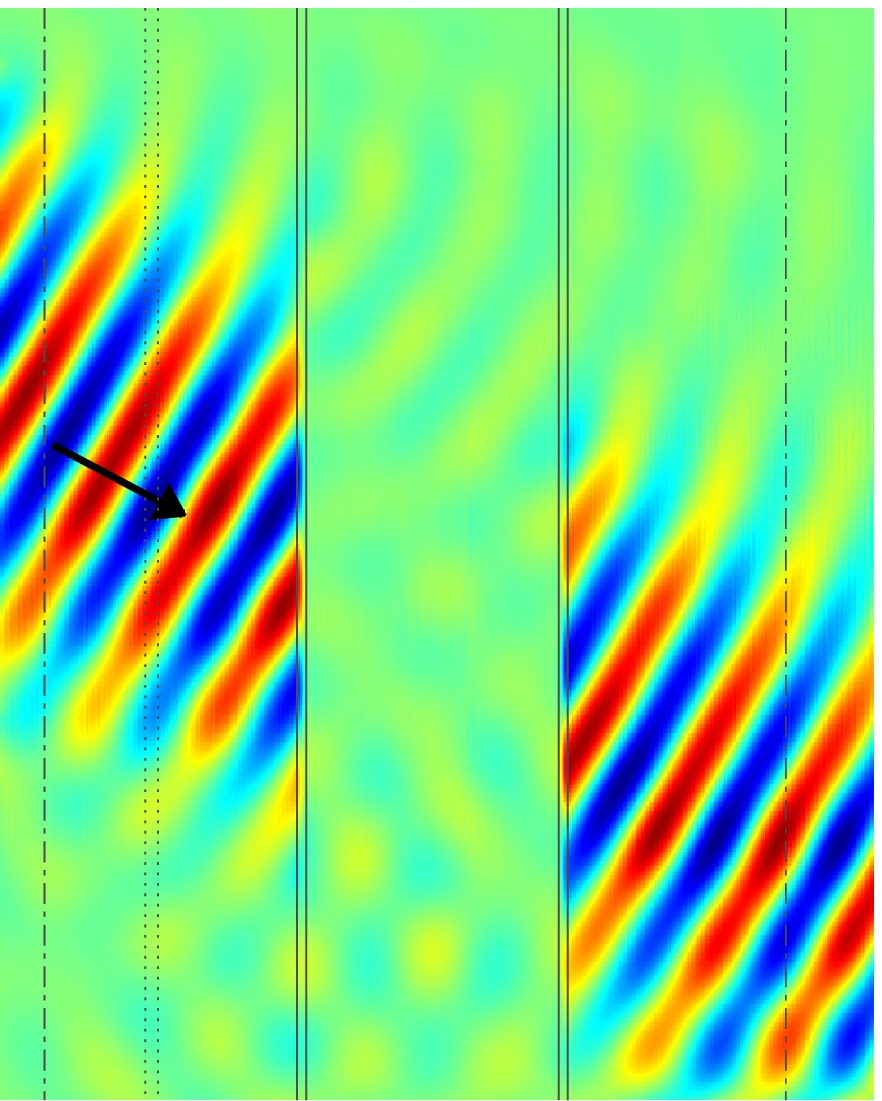}  &  \includegraphics[height=4.9 cm, valign=c]{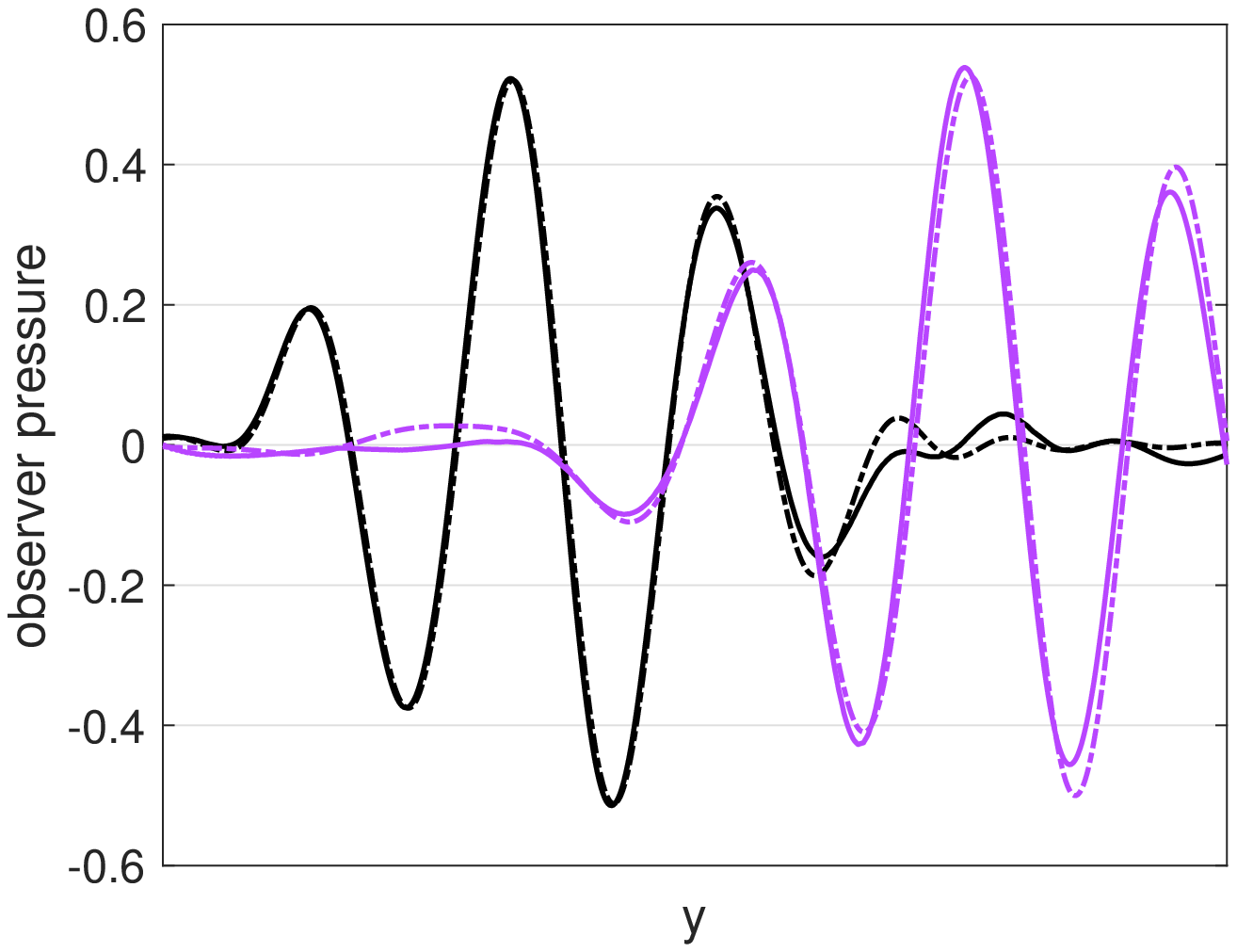}
\end{tabular} 
\includegraphics[width=4.0 cm]{Figures/colorbar_jet.eps}
\end{center}
 \caption{Cloak performance demonstration for a varying incidence angle $\theta$. Plotted are snapshots of the pressure field time responses to a detection beam emitted from the left boundary of a $2.4\times 3.0$ m$^2$ air waveguide with $\epsilon=0.025$ m. Rows 1:4 correspond to source angles $\theta=30^{\textrm{o}}, 15^{\textrm{o}}, 0^{\textrm{o}}, -30^{\textrm{o}}$. Left column: entire waveguide responses when the cloak is off, giving free space propagation (black arrow - propagation direction). Middle column: entire waveguide responses when the cloak is on, resulting in the required dead zone formation (vertical lines the same as in Fig. \ref{lambda_change}). Right column: comparison of cloak-on (solid) and cloak-off (dashed-dotted) responses at the left (black) and right (purple) observers.}
\label{theta_change}
\end{figure*}

We now demonstrate the dynamical performance of the cloaking system of Fig. \ref{GenScheme} according to the control algorithm derived in Sec. \ref{Algorithm}. We choose the waveguide medium to be air, implying mass density $\rho_0=1.2$ kg/m$^3$ and bulk modulus $b_0=1.42\cdot 10^5$ N/m$^2$. 
The numerical experiments are carried out via a $2D$ finite difference time domain procedure with spatial and temporal discretization steps of $a=0.005$ and $dt=10^{-6}$, respectively.
The overall waveguide dimensions are set to $L_x\times L_y=2.4\times 3.0$ m$^2$. Absorbing boundary conditions are modeled along the waveguide boundaries. 
The locations of the $f_o$ and $f_c$ control input arrays, the two sensor arrays, and the left and right observer arrays, are respectively given by $x_o=0.35$, $x_c=0.75$, $x_{m1}=0.39$, $x_{m2}=0.43$, $O_l=0.12$ and $O_r=2.16$ m. \\
At the first step we examine the trade-off between the ratio $\lambda/\epsilon$, wavelength over actuation arrays spacing, and the control effort. This trade-off was analyzed in Sec. \ref{uni}, where it was shown that the control input that is traded off is $h$, which is responsible for the uni-directionality of the total control wave.
The trade-off is illustrated in Fig. \ref{trade_off}. Subplot (a) depicts the frequency response diagrams of the control input amplitude $|C_h(i\omega)|$ (dashed-dotted lines) versus the backwave amplitude $|\overline{Q}(i\omega)|$ (solid lines), defined in \eqref{eq:C_h} and \eqref{eq:Q}, for different values of $\epsilon$.
The blue and purple curves respectively correspond to $\epsilon=0.02$ m and $\epsilon=0.1$ m, whereas the black curves correspond to $\epsilon=0.05$ m (solid) and $\epsilon=0.025$ m (dotted). 
To relate to a response with a fixed ratio $\lambda/\epsilon$, we consider two detection frequencies, $\omega_{\psi1}=5.4\cdot 10^3$ r/s (or $0.86$ kHz), implying $\lambda=0.4$ m, and $\omega_{\psi2}=10.8\cdot 10^3$ r/s (or $1.72$ kHz), implying $\lambda=0.2$ m. This results in $\lambda/\epsilon=20,8,4$ for the blue, black and purple curves at the corresponding frequencies, respectively.
As was predicted in Sec. \ref{uni}, the bigger is the control effort, the smaller is the backwave. Here, for $\omega_{\psi1}$, backwave amplitudes of $-14$ dB, $-7$ dB and $-2$ dB require a control effort of $10$ dB, $3$ dB and $-4$ dB.
The actual values of $\lambda$ and $\epsilon$ nearly do not affect the amplitudes as long as their ratio is preserved, as demonstrated for $\lambda/\epsilon=8$ at $\omega_{\psi1}$ and $\omega_{\psi2}$. \\
Subplots (b)-(e) depict steady-state snapshots of the pressure field time responses to the total control wave at $x_o$ and $x_o-\epsilon$ according to \eqref{eq:v_fh_lim} for $\theta=22^\textrm{o}$. 
The command for $f_o$ is a harmonic signal $\psi(y,t)$ of frequency $\omega$ and a geometrical phase distribution $\Delta\varphi(y)$. $\omega$ and $\Delta\varphi(y)$ uniquely determine the resulting wavelength $\lambda=2\pi c/\omega$ and incidence angle $\theta$, respectively.
Snapshots (b),(c),(e) correspond to the fixed wavelength $\lambda=0.4$ m, and to the three distances $\epsilon=0.02,0.05$ and $0.1$ m, whereas snapshot (d) corresponds to $\lambda=0.2$ m and $\epsilon=0.025$ m, in accordance with Fig. \ref{trade_off}(a).
The main control wave propagates in $x>x_o$, whereas the backwave propagates in $x<x_o$. As expected, the backwave increases as $\lambda/\epsilon$ decreases. 
To complete the picture, subplot (f) depicts the $y$ axis cross-section of the $h_o$ component of the responses in Fig. \ref{trade_off}(b)-(e) along the actuation section $d$ at times instances when the control signal amplitude is the highest. For $\lambda/\epsilon=20,8$ and $4$ with $\lambda=0.4$ m the maximal amplitudes are in accordance with the trade-off prediction of the frequency response in subplot (a). 
For $\lambda/\epsilon=8$ with $\lambda=0.2$ m the spatial distribution of $h_o(y,t)$ is different but the maximal amplitudes ratio remains in the same order of magnitude. \\
The next step is demonstrating the performance of the actual cloak, which is the dead zone creation process. In the following we consider a fixed actuators $x$ axis spacing of $\epsilon=0.025$ m, where all the adaptation to the changing detection properties is carried out in real-time through the controllers $C_\psi$, $C_h$ and $C_{fo,c}$, designed in Sec. \ref{Algorithm}. 
In Fig. \ref{lambda_change} we test our algorithm under frequency variation. We launch a beam in four time intervals, having a fixed incidence angle $\theta=22^{\textrm{o}}$ in all the intervals, and a different wavelength in each: $\lambda=0.1$ m, first row, $\lambda=0.2$ m, second row, $\lambda=0.3$ m, third row, and $\lambda=0.4$ m, fourth row. These wavelengths respectively correspond to the frequencies $\omega=3.4$ kHz, $\omega=1.72$ kHz, $\omega=1.14$ kHz and $\omega=0.86$ kHz, resulting in the ratio $\lambda/\epsilon$ of $4$, $8$, $12$ and $16$. The four beams are launched with the same amplitude $A=1$, the same vertical shift of $y=0.45$ m, but have a different $y$ axis extension $d$ and a different Gaussian scaling. \\
The left column depicts steady-state time snapshots of the pressure field responses in the waveguide when the cloaking system is turned off, giving a free field propagation. 
The middle column depicts the corresponding responses when the cloaking system is turned on, resulting in a dead zone between the two pairs of actuator arrays, at $x_o,x_o-\epsilon$ and $x_c,x_c-\epsilon$. These arrays are pictured by solid vertical lines, the measurement arrays at $x_{m1},x_{m2}$ by dotted lines, and the observer arrays at $O_l,O_r$ by dashed-dotted lines. In all the intervals the algorithm managed to generate the required opening and closing control beams in real-time. The smaller is the wavelength, the higher is the resulting backwave amplitude, as seen from the field intensity maps of the snapshots. The minimal pressure reduction in the dead zone was by 10 times of the original detection signal, obtained for $\lambda=0.1$ m. The maximal reduction was by 20 times, obtained for $\lambda=0.4$ m.
The right column depicts the $y$ cross-section of the pressure field time responses in the left and middle columns at the detection observers arrays. The figures compare the responses along the left observer at $x=O_l$, black, and right observer at $x=O_r$, purple, at the cloak-off (free space) state, dashed, and the cloak-on state (dead zone), solid. The dashed and solid responses are reasonably close. The largest deviation occurs at the right observer at the backwave region. \\
In Fig. \ref{theta_change} we test our algorithm under incidence angle variation. We launch a detection beam of a fixed wavelength $\lambda=0.2$ m (frequency $\omega=1.72$ kHz) during four time intervals, in which the incidence angle $\theta$ is switched through $30^\textrm{o}$, $15^\textrm{o}$, $0^\textrm{o}$ and $-30^\textrm{o}$. The $y$ axis beam extension, its amplitude and Gaussian scaling are identical in all the intervals.
The four values of $\theta$ respectively correspond to the first, second, third and fourth rows of the table.
Similarly to Fig. \ref{lambda_change}, here the left column depicts free space 2D pressure field response, the middle column depicts the response when the cloak is turned on, and the right column depicts the comparison between the two (dashed-dotted vs. solid), as observed by the detection system at $O_l$, black, and $O_r$, purple.
The pressure field intensity average inside the dead zone is identical for the four intervals, given by a 20 times reduction of the source wave.

\section{Conclusion}   \label{Conclusion}

We studied the problem of active acoustic cloaking of objects that can move in a two-dimensional medium. 
The cloaking goal was to create a real-time reconfigurable dead zone, which is an artificial quiet channel for the object to pass through undetected,
under several constraints that may arise in practical applications. Specifically, we assumed that (i) the detection system emits signals from one side of the object but observers the response on both of its sides, (ii) the detection signal is a series of harmonic beams launched at finite time intervals, and have their properties, such as frequency, phase, amplitude, distance from object, incidence angle etc., changed between the intervals, (iii) the control actuators and sensors cannot block the medium, i.e. cannot constitute a boundary themselves, (iv) the object is non-emitting but of an unknown and possibly varying impedance, and (v) the object cannot be covered by a rigid shell. \\
We proposed a solution that we denoted by a mid-domain near uni-directional wave generation. 
This technique enables to launch control beams in a desired direction in the domain interior that can be steered around with a minimal back action wave as a trade-off with the control effort. 
We demonstrated the idea in a two-dimensional waveguide platform with gapped plates, depicted in Fig. \ref{GenScheme}. The cloak was executed by two pairs of monopole actuator arrays and one pair of sensor arrays, which were mounted in the interior of one of the plates, facing inwards but not blocking the gap. 
Our control algorithm for the actuator inputs, given by \eqref{eq:C_h}-\eqref{eq:f_final}, generated two near uni-directional beams.
One beam opened the dead zone by intercepting the detection signal before it hit the object, thus preventing its coupling with the object impedance.
The second beam closed the dead zone by reconstructing the original signal and imitating free space propagation for the observers. 
The time evolution of both control beams was based on real-time pressure field prediction at the actuation locations, based on information assumed to be available from measurements, according to the detection intervals that are illustrated in Fig. \ref{fig:estimation}. \\
We tested our cloak performance via time domain numerical simulations of a $2.4\times 3$ m$^2$ waveguide filled with air.
First, in Fig. \ref{trade_off} we demonstrated the trade-off between the level of control beam uni-directionality, exhibited by the backwave amplitude minimization, and the control effort required to achieve it for a given actuator arrays spacing.
Then, keeping the actuators locations fixed, we changed a certain property of the detection beam. In Fig. \ref{lambda_change} we imitated detection sources of a varying frequency by launching a beam at an angle $\theta=22^\textrm{o}$ with wavelength switching through $\lambda=0.1,0.2,0.3,0.4$ m at each new time interval. 
The longer was the wavelength the smaller was the backwave, resulting in up to 20 times of amplitude reduction in the dead zone.
In Fig. \ref{theta_change} we imitated steering detection beams by scanning the area at angles switching through $\theta=30^\textrm{o},15^\textrm{o},0^\textrm{o},-30^\textrm{o}$. The amplitude reduction in the dead zone was about 20 times for all the angles.

\section*{Acknowledgement}

We thank Viacheslav (Slava) Krylov and Amir Boag for insightful discussions.


\bibliographystyle{IEEEtran}        
\bibliography{AcousticAbsorberCloak}           



\end{document}